\documentclass[showpacs,aps,graphicx,twocolumn,]{revtex4}
\usepackage{amsmath}
\usepackage{mathrsfs}
\usepackage{amsfonts}
\usepackage{amssymb}
\usepackage{graphicx}
\usepackage{eufrak}
\usepackage{multirow}
\usepackage{float}
\begin{document}

\title{Low-cost Fredkin gate with auxiliary space}

\author{Wen-Qiang Liu$^{1}$, Hai-Rui Wei$^{1}$\footnote{Corresponding author: hrwei@ustb.edu.cn} and Leong-Chuan Kwek$^{2,3,4}$}

\address{$^1$ School of Mathematics and Physics, University of Science and Technology Beijing, Beijing 100083, China\\
        $^2$ Centre for Quantum Technologies, National University of Singapore, Singapore 117543, Singapore \\
        $^3$ MajuLab, CNRS-UNS-NUS-NTU International Joint Research Unit, Singapore UMI 3654, Singapore\\
        $^4$ National Institute of Education and Institute of Advanced Studies, Nanyang Technological University, Singapore 637616, Singapore}

\date{\today }


\begin{abstract}
Effective quantum information processing is tantamount in part to the minimization the quantum resources needed by quantum logic gates. Here, we propose an optimization of an $n$-controlled-qubit Fredkin gate with a maximum of $2n+1$ two-qubit gates and $2n$ single-qudit gates by exploiting auxiliary Hilbert spaces. The number of logic gates required improves on earlier results on simulating arbitrary $n$-qubit Fredkin gates. In particular, the optimal result for one-controlled-qubit Fredkin gate (which requires three qutrit-qubit  partial-swap gates) breaks the theoretical nonconstructive lower bound of five two-qubit gates. Furthermore, using an additional spatial-mode degree of freedom, we design a possible architecture to implement a polarization-encoded Fredkin gate with linear optical elements.
\end{abstract}

\pacs{03.67.Lx, 42.50.Ex, 42.79.Ta} \maketitle


\section{Introduction}\label{sec1}


The realization of a fully scalable universal programmable quantum computer with currently available technology, albeit promising, remains a major challenge in the field of quantum information science \cite{book}. Similarly to classical computer architectures, elementary logic gates provide the Lego blocks for the building of large-scale quantum computers \cite{Barenco}. For universal quantum computing, it has been shown that two-qubit entangling gates and arbitrary single-qubit rotations are sufficient for realizing any quantum circuit \cite{Barenco}. Quantum logic gates have been experimentally demonstrated using ion traps \cite{ions1,ions2}, nuclear magnetic resonance \cite{NMR}, integrated optics \cite{Integrated-optics1,Integrated-optics2}, photons \cite{atom-based1,atom-based2,multiphoton,frequency}, superconductors \cite{superconducting1,superconducting2}, atom-photon systems \cite{hybrid}, atoms \cite{atom1,atom2}, quantum dots \cite{QD1}, and nitrogen-vacancy defect centers \cite{NV}. Yet, one of the main difficulties that hovers over the true application of quantum computation in the real world is the control and manipulation of a large number of elementary gates in realistic experiments with a computational complexity (also called the cost, measured by the number of two-qubit entangling gates needed to perform the computation) that grows exponentially  with the number of qubits. Therefore, a more effective method is required for building universal quantum circuits with less resource overhead and a minimum of external manipulation.

The implementation of elementary multiqubit gates is an important milestone on the way to a scalable quantum computer. Multiqubit conditional gates such as Fredkin (controlled-swap) and Toffoli (controlled-controlled-NOT) gates are crucial for certain practical quantum information-processing applications ranging from quantum error correction \cite{error-correction} and quantum fault tolerance \cite{Fault-tolerant} to quantum algorithms \cite{algorithms} and quantum entangling operations \cite{entangling-operations}.
A Fredkin or Toffoli gate is universal for multiqubit quantum computing when combined with single-qubit Hardmard gates \cite{Fredkin1,Fredkin2}.
In 1995, Chau and Wilczek \cite{Fredkin6} showed that a Fredkin gate can be built using six two-body reversible gates.
In 1996, Smolin and DiVincenzo \cite{Fredkin5-1} constructed a Fredkin gate with five two-qubit gates.
Later, in 2015, Ivanov \emph{et al.} \cite{global} presented an improved synthesis of a Fredkin gate using four global or five nearest-neighbor gates.
Yu and Ying \cite{Fredkin5-2}  demonstrated theoretically that at least five two-qubit entangling gates are required to simulate a three-qubit Fredkin gate.
If we restrict our attention to controlled-NOT (CNOT) gates, the optimal cost of a Fredkin gate is eight CNOT gates because of $U_{\text{Fredkin}}=I_2\otimes U_{\text{CNOT}} \cdot U_{\text{Toffoli}}\cdot I_2\otimes U_{\text{CNOT}}$, where $I_2$ is the $2\times2$ identity matrix \cite{Fredkin-optimal1,Frekin-optimal2}.
Notably, the number of gates required for the construction of quantum circuits can be further optimized with the aid of auxiliary dimensions or
degrees of freedom (DOFs) \cite{T-PRA,T-NatPhy,Multivalue,multilevel4}.
By using multilevel information carriers, Ralph \emph{et al}. \cite{T-PRA} and  Lanyon \emph{et al}. \cite{T-NatPhy}  reduced the cost of a Toffoli gate from six CNOT gates \cite{Fredkin-optimal1} to three CNOT gates, and such results beat the theoretical lower bound of five two-qubit entangling gates \cite{lower-T}. Liu and Wei \cite{Liu} also reduced the complexity of a Fredkin gate from eight CNOT gates \cite{Fredkin-optimal1,Frekin-optimal2} to five CNOT gates assisted by auxiliary dimensions.
In other words, the architecture of a Fredkin gate might further break the theoretical lower bound of five \cite{Fredkin5-2} by using auxiliary dimensions or DOFs.

Utilizing linear optics, an auxiliary entangled pair of photons, and two-qubit and single-qubit gates, a variety of decomposition-based three-qubit Fredkin gates have been proposed \cite{construction1,Gong,construction2,Ono}. A controlled-phase-flip-based linear-optics Fredkin gate with a maximum success probability of $4.1\times10^{-3}$ was proposed by Fiur\'{a}\v{s}ek \cite{construction1}  in 2006. In 2008, Gong \emph{et al}. \cite{Gong} presented a postselected scheme for implementing a CNOT-based Fredkin gate with a higher success probability of $1/192$. Subsequently, Fiur\'{a}\v{s}ek \cite{construction2} further increased the success probability to $1/162$ with a coincidence basis, and such a scheme was experimentally realized with a fidelity of $0.85\pm0.03$ by Ono \emph{et al.} \cite{Ono} in 2017.
In an alternative approach, by exploiting four-photon path-mode entanglement, Patel \emph{et al.} \cite{Patel} experimentally realized a polarization Fredkin gate without decomposition  in 2016. Experiments with multiple DOFs have also been done. In 2018, a two-photon three-qubit polarization-spatial Fredkin gate was experimentally demonstrated by  St\'{a}rek \emph{et al.} \cite{polarization-saptial}  with a success probability of 1/9. There was also an experiment by Urrego \emph{et al.} in 2020 \cite{polarization-OAM} using polarization and orbital angular momentum.



In this study, we further reduce the cost of a three-qubit Fredkin gate from five two-qubit gates to three fundamental qutrit-qubit gates (referred to as partial-swap gates) by introducing auxiliary Hilbert spaces (with three levels). We expand the qutrit (three-level) design to a qudit design (more than three levels) and propose a low-cost quantum circuit for implementing an $n$-controlled-qubit Fredkin gate that requires only $2n+1$ two-qubit gates and $2n$ single-qudit gates. In addition, using linear optical elements and a single-photon source, we construct a postselected partial-swap gate with a success probability of $1/2$ without the assistance of any auxiliary photons. Based on our proposed gate circuit and the linear optical architecture of the partial-swap gate, a low-cost photonic Fredkin gate is thus achieved.

Our schemes have the following favorable features: (a) the cost of a one-controlled-qubit Fredkin gate is three partial-swap gates, which beats a five-CNOT-gate construction with a higher-dimensional ancilla \cite{T-NatPhy,T-PRA}; (b) a specific detailed compact Fredkin gate circuit is designed; (c) the $n$-controlled-qubit Fredkin gate requires only $2n+1$ two-qubit gates, and therefore it beats the previous proposals that require $O(n^2)$ \cite{Barenco} or $2n+3$ gates \cite{T-NatPhy}; (d) the success probability of our partial-swap gate, 1/2, is better than that of a postselected CNOT gate, $1/9$ \cite{CNOT-BS1,CNOT-BS2,CNOT-BS3,CNOT-BS4}; and (e) optical realizations of our scheme are possible using existing technology. Moreover, our proposed resource-saving techniques can potentially contribute to the optimization of larger-scale quantum computational networks.

\section{Simplifying Fredkin circuits using higher energy levels} \label{sec2}

\subsection{Synthesis of a three-qubit Fredkin gate using qutrits} \label{sec2.1}

\begin{figure} [htpb]
\begin{center}
\includegraphics[width=6 cm,angle=0]{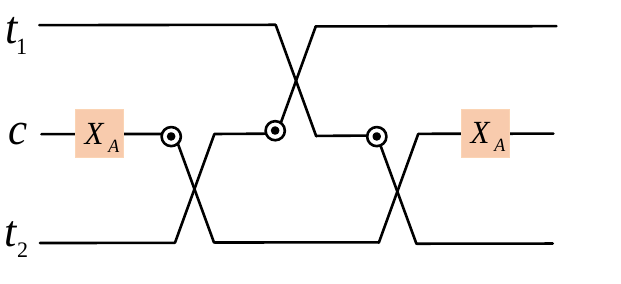}
\caption{(Color online) Simplified synthesis of a three-qubit Fredkin gate. The single-qutrit gate $X_A$ completes the transformation $|0\rangle \leftrightarrow |2\rangle$.
The circles $\odot$, coded with $|0\rangle$ and $|1\rangle$ entries, are used to define the action position of a partial-swap gate, i.e., a swap gate is encountered if and only if the input state of $c$ is $|0\rangle$ or $|1\rangle$.} \label{3-qubit-synthesis}
\end{center}
\end{figure}

Our proposal for the implementation of the simplified three-qubit Fredkin gate is shown in Fig. \ref{3-qubit-synthesis}. We present the details of our method according to the following steps. We assume that the system is initially prepared in an arbitrary state as
\begin{eqnarray}              \label{eq1}
\begin{split}
|\phi_0\rangle=&\alpha_1|0\rangle_c|0\rangle_{t_1}|0\rangle_{t_2}+\alpha_2|0\rangle_c|0\rangle_{t_1}|1\rangle_{t_2}\\
              &+\alpha_3|0\rangle_c|1\rangle_{t_1}|0\rangle_{t_2}+\alpha_4|0\rangle_c|1\rangle_{t_1}|1\rangle_{t_2}\\
              &+\alpha_5|1\rangle_c|0\rangle_{t_1}|0\rangle_{t_2}+\alpha_6|1\rangle_c|0\rangle_{t_1}|1\rangle_{t_2}\\
              &+\alpha_7|1\rangle_c|1\rangle_{t_1}|0\rangle_{t_2}+\alpha_8|1\rangle_c|1\rangle_{t_1}|1\rangle_{t_2},
\end{split}
\end{eqnarray}
where the coefficients $\alpha_i$ ($i=1, 2, \cdots, 8$) are complex amplitudes and satisfy the condition that $\sum^{8}_{i=1} |\alpha_{i}|^{2}=1$. The subscripts $c$, $t_1$, and $t_2$ denote the control, first target, and second target qubits, respectively.

First, the control qubit $c$ encounters a single-qutrit gate $X_A$, which moves the computational state $|0\rangle$ to the third-level auxiliary state $|2\rangle$, thereby resulting in
\begin{eqnarray}              \label{eq2}
\begin{split}
|\phi_1\rangle=&\alpha_1|2\rangle_c|0\rangle_{t_1}|0\rangle_{t_2}+\alpha_2|2\rangle_c|0\rangle_{t_1}|1\rangle_{t_2}\\
              &+\alpha_3|2\rangle_c|1\rangle_{t_1}|0\rangle_{t_2}+\alpha_4|2\rangle_c|1\rangle_{t_1}|1\rangle_{t_2}\\
              &+\alpha_5|1\rangle_c|0\rangle_{t_1}|0\rangle_{t_2}+\alpha_6|1\rangle_c|0\rangle_{t_1}|1\rangle_{t_2}\\
              &+\alpha_7|1\rangle_c|1\rangle_{t_1}|0\rangle_{t_2}+\alpha_8|1\rangle_c|1\rangle_{t_1}|1\rangle_{t_2}.
\end{split}
\end{eqnarray}

Next, three qutrit-qubit partial-swap gates are applied to $c$ and $t_2$, $c$ and $t_1$, and $c$ and $t_2$ in succession, and these operations exchange information between $c$ and $t_2$ (and between $c$ and $t_1$) depending on whether $c$ is a 0 or a 1 entry, i.e., the partial-swap (p-swap) gate performs the transformations
\begin{eqnarray}              \label{eq3}
\begin{split}
&|00\rangle \xrightarrow{\text{p-swap}} |00\rangle, \quad &|01\rangle \xrightarrow{\text{p-swap}} |10\rangle, \\
&|10\rangle \xrightarrow{\text{p-swap}} |01\rangle, \quad &|11\rangle \xrightarrow{\text{p-swap}} |11\rangle, \\
&|20\rangle \xrightarrow{\text{p-swap}} |20\rangle, \quad &|21\rangle \xrightarrow{\text{p-swap}} |21\rangle.
\end{split}
\end{eqnarray}
Therefore, the three partial-swap gates change $|\phi_1\rangle$ to
\begin{eqnarray}              \label{eq4}
\begin{split}
|\phi_2\rangle=&\alpha_1|2\rangle_c|0\rangle_{t_1}|0\rangle_{t_2}+\alpha_2|2\rangle_c|0\rangle_{t_1}|1\rangle_{t_2}\\
              &+\alpha_3|2\rangle_c|1\rangle_{t_1}|0\rangle_{t_2}+\alpha_4|2\rangle_c|1\rangle_{t_1}|1\rangle_{t_2}\\
              &+\alpha_5|1\rangle_c|0\rangle_{t_1}|0\rangle_{t_2}+\alpha_6|1\rangle_c|1\rangle_{t_1}|0\rangle_{t_2}\\
              &+\alpha_7|1\rangle_c|0\rangle_{t_1}|1\rangle_{t_2}+\alpha_8|1\rangle_c|1\rangle_{t_1}|1\rangle_{t_2}.
\end{split}
\end{eqnarray}

Finally, the $X_A$ gate is applied again to reshape the original computational qubits (the state $|2\rangle$ returns back to $|0\rangle$), thereby yielding
\begin{eqnarray}              \label{eq5}
\begin{split}
|\phi_3\rangle=&\alpha_1|0\rangle_c|0\rangle_{t_1}|0\rangle_{t_2}+\alpha_2|0\rangle_c|0\rangle_{t_1}|1\rangle_{t_2}\\
              &+\alpha_3|0\rangle_c|1\rangle_{t_1}|0\rangle_{t_2}+\alpha_4|0\rangle_c|1\rangle_{t_1}|1\rangle_{t_2}\\
              &+\alpha_5|1\rangle_c|0\rangle_{t_1}|0\rangle_{t_2}+\alpha_6|1\rangle_c|1\rangle_{t_1}|0\rangle_{t_2}\\
              &+\alpha_7|1\rangle_c|0\rangle_{t_1}|1\rangle_{t_2}+\alpha_8|1\rangle_c|1\rangle_{t_1}|1\rangle_{t_2}.
\end{split}
\end{eqnarray}

Based on Eqs. (\ref{eq1})--(\ref{eq5}), we find that the three partial-swap gates are sufficient to implement a three-qubit Fredkin gate, and this realization breaks the theoretical lower bound of five two-qubit gates \cite{Fredkin5-2}. The key optimization technique involves a single-qutrit operator $X_{A}$ that temporarily provides the Hilbert subspace for the control qubit, and the $X_{A}$ operator completes the transformations $X_A|0\rangle=|2\rangle$, $X_A|2\rangle=|0\rangle$, and $X_A|1\rangle=|1\rangle$.

\begin{figure*}  [htbp]
\begin{center}
\includegraphics[width=11.0 cm,angle=0]{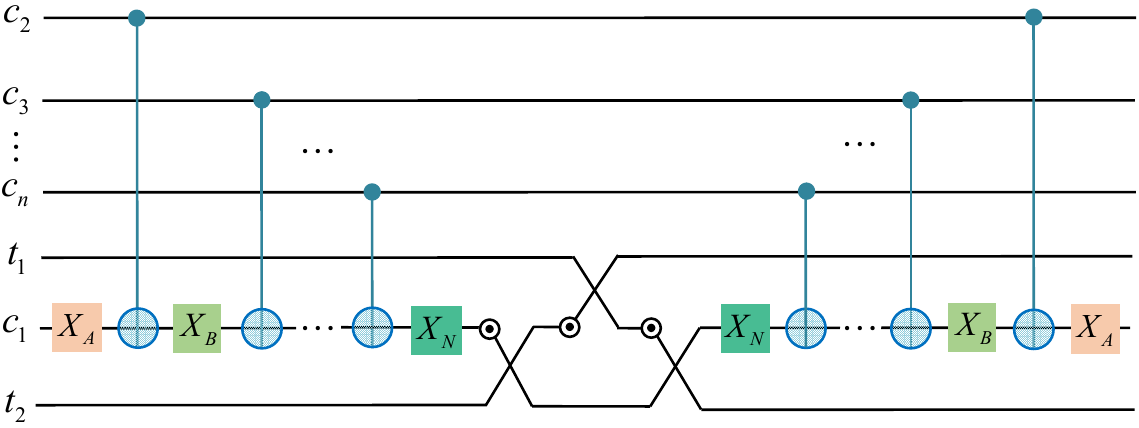}
\caption{(Color online) Simplified synthesis of an $n$-controlled-qubit Fredkin gate with $2n+1$ two-qubit gates and $2n$ single-qudit gates. Single-qudit gates $X_A$, $X_B$, $\cdots$, $X_N$ complete the transformations $|0\rangle \leftrightarrow |2\rangle$, $|1\rangle \leftrightarrow |3\rangle$, $\cdots$, $|0\rangle \leftrightarrow |n+1\rangle$ when $n$ is odd or $|0\rangle \leftrightarrow |2\rangle$, $|1\rangle \leftrightarrow |3\rangle$, $\cdots$, $|1\rangle \leftrightarrow |n+1\rangle$ when $n$ is even. All CNOT and partial-swap gates perform at the qubit-level. The control $\bullet$ ($\odot$) turns on for the input of $|1\rangle$ ($|0\rangle$ or $|1\rangle$). } \label{n-controlled-synthesis}
\end{center}
\end{figure*}

\subsection{Extension to multiple-controlled-qubit Fredkin gate circuits} \label{sec2.2}


In the previous subsection, we construct Fredkin gates for the three-qubit case. However, our optimization technique is generic and   is also suitable for the multiple-qubit case (i.e., for a system with more than three qubits). Inspired by the work of Lanyon \emph{et al}. \cite{T-NatPhy}, by introducing $n$ extra auxiliary levels into the first control-qubit carrier, we construct an $n$-controlled-qubit Fredkin gate with $2(n-1)$ CNOT gates, three partial-swap gates, and $2n$ single-qudit operators. A program for simulating an $n$-controlled-$U_k$ gate with $2n-2$ CNOT gates and one controlled-$U_k$ gate was proposed by Lanyon \emph{et al}. \cite{T-NatPhy}, but a construction of the controlled-$U_k$ gate is still an open question. Here, $U_k$ is an arbitrary unitary operation acting on $k$ qubits. The result of Ref. \cite{T-NatPhy} indicates that $2n+3$ CNOT gates are sufficient to simulate an $n$-controlled-qubit Fredkin gate.

As shown in Fig. \ref{n-controlled-synthesis}, the ($n+2$)-dimensional Hilbert space is established by using single-qudit operations comprising $X_A$, $X_B$, $\cdots$, $X_N$ for the synthesis of the $n$-controlled-qubit Fredkin gate, where $X_A$, $X_B$, $\cdots$, $X_N$ implement the transformations  $|0\rangle \leftrightarrow |2\rangle$, $|1\rangle \leftrightarrow |3\rangle$, $|0\rangle \leftrightarrow |4\rangle$, $\cdots$, $|1\rangle \leftrightarrow |n\rangle$, $|0\rangle \leftrightarrow |n+1\rangle$ when $n$ is odd or $|0\rangle \leftrightarrow |2\rangle$, $|1\rangle \leftrightarrow |3\rangle$, $|0\rangle \leftrightarrow |4\rangle$, $\cdots$,  $|0\rangle \leftrightarrow |n\rangle$, $|1\rangle \leftrightarrow |n+1\rangle$ when $n$ is even.
%


\section{Linear optical Fredkin gates for polarization-encoded qubits} \label{sec3}

\begin{figure}[htpb]
\begin{center}
\includegraphics[width=8.6 cm,angle=0]{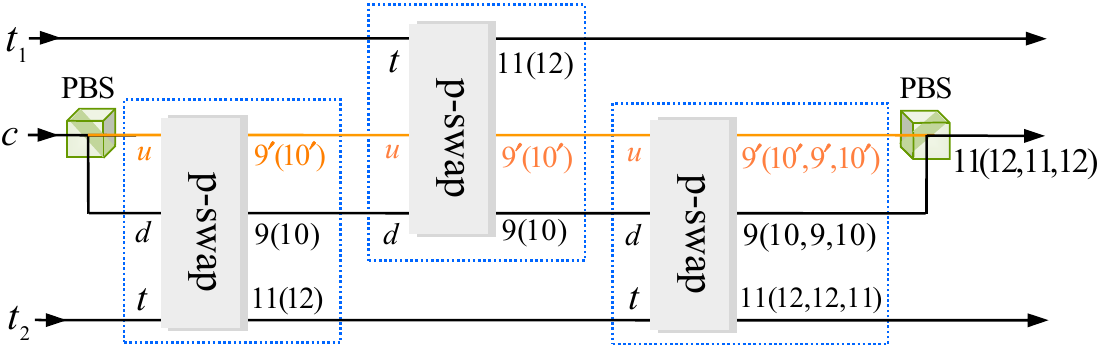}
\caption{(Color online) Schematic illustration of the three-qubit Fredkin gate. ``PBS'' represents a polarizing beam splitter, which transmits an $H$-polarized photon and reflects a $V$-polarized photon. $u$, $d$, and $t$ are the input ports. 9, $9'$, 10, $10'$, 11, and 12 are the output ports.} \label{post-selected-1}
\end{center}
\end{figure}


We now consider a photonic implementation of the three-qubit Fredkin gate with linear optics. The computational basis is encoded in polarization states $|H\rangle=|0\rangle$ and $|V\rangle=|1\rangle$, where $|H\rangle$ and $|V\rangle$ are  horizontally and vertically polarized photons, respectively. As shown in Fig. \ref{post-selected-1},
polarizing beam splitters (PBSs) act as single-qutrit $X_A$ gates by adding an extra spatial mode such that $|2\rangle=|H_u\rangle$ (brown), $|1\rangle=|V_d\rangle$ (black), and $|0\rangle=|H_d\rangle$ (black). Henceforth, $|H_i\rangle$ ($|V_i\rangle$) represents the $H$- ($V$-) polarized component emitted from mode $i$.  Our next task involves the construction of another key component, comprising the partial-swap gate, which does not exchange the states of the two photons if the first photon is in the state where $|2\rangle=|H_{u}\rangle$.

\begin{figure} [!h]    
\begin{center}
\includegraphics[width=8.4 cm,angle=0]{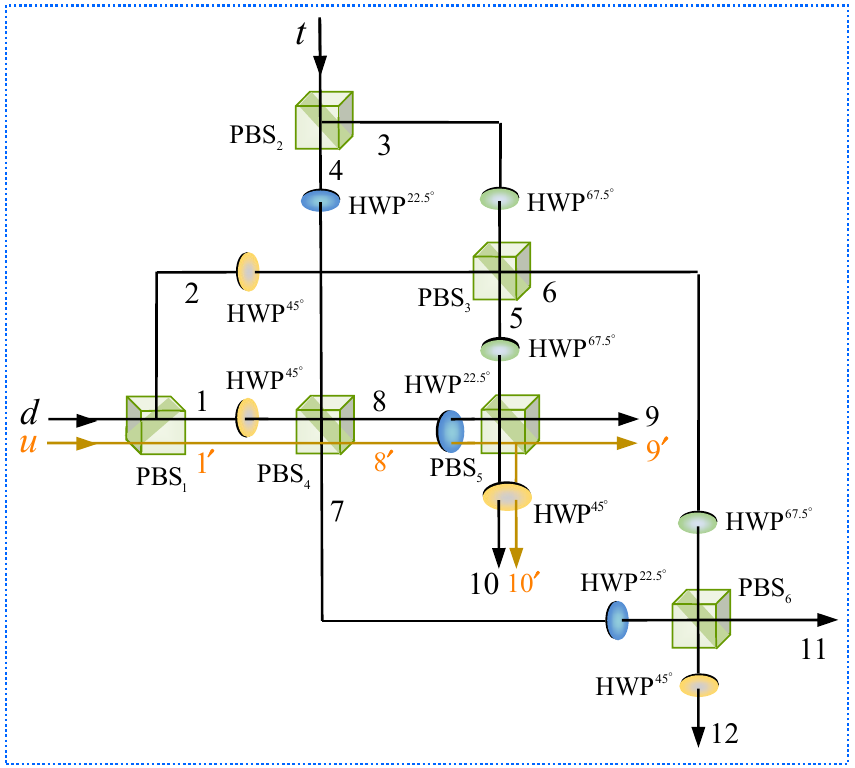}
\caption{(Color online) Schematic illustration of the postselected partial-swap gate. HWP$^{45^\circ}$ represents a half-wave plate (HWP) oriented at 45$^\circ$, resulting in $|H\rangle\leftrightarrow|V\rangle$. HWP$^{22.5^\circ}$ (HWP$^{67.5^\circ}$), set to 22.5$^\circ$ (67.5$^\circ$), induces transformations
$|H\rangle\leftrightarrow(|H\rangle+|V\rangle)/\sqrt{2}$ and
$|V\rangle\leftrightarrow(|H\rangle-|V\rangle)/\sqrt{2}$
 ($|H\rangle\leftrightarrow(-|H\rangle+|V\rangle)/\sqrt{2}$ and
$|V\rangle\leftrightarrow(|H\rangle+|V\rangle)/\sqrt{2}$).} \label{post-selected-2}
\end{center}
\end{figure}

The proposed setup for implementing a linear optical partial-swap gate is shown in Fig. \ref{post-selected-2}. We assume that the state of the system is prepared in the general initial state
\begin{eqnarray}              \label{eq6}
\begin{split}
|\psi_0\rangle=&
 \alpha_1|H_u\rangle|H\rangle
+\alpha_2|H_u\rangle|V\rangle
+\alpha_3|H_d\rangle|H\rangle\\&
+\alpha_4|H_d\rangle|V\rangle
+\alpha_5|V_d\rangle|H\rangle
+\alpha_6|V_d\rangle|V\rangle.
\end{split}
\end{eqnarray}

First, the polarizing beam splitter PBS$_1$ transmits $H_u$ (brown, emitted from mode $u$) and $H_d$ (black, emitted from mode $d$) for the first photon into mode $1'$ and mode 1, and reflects $V_d$ (black, emitted from mode $d$) for the first photon into mode 2.
PBS$_2$ then transmits (reflects) the $H$- ($V$-) polarized component of the second photon from mode $t$ into mode $4$ (mode 3). Next, the photons in modes 1, 2, 3, and 4 pass through HWP$^{45^\circ}$, HWP$^{45^\circ}$, HWP$^{67.5^\circ}$, and HWP$^{22.5^\circ}$, respectively. Here, HWP$^{45^\circ}$ represents a half-wave plate oriented at $45^\circ$, which induces a bit-flip operation $|H\rangle \leftrightarrow |V\rangle$. In addition, HWP$^{22.5^\circ}$, set to $22.5^\circ$, induces
\begin{eqnarray}                  \label{eq7}
\begin{split}
&|H\rangle\leftrightarrow\frac{1}{\sqrt{2}}(|H\rangle+|V\rangle), \quad
|V\rangle\leftrightarrow\frac{1}{\sqrt{2}}(|H\rangle-|V\rangle).
\end{split}
\end{eqnarray}
HWP$^{67.5^\circ}$, set to $67.5^\circ$, results in
\begin{eqnarray}                  \label{eq8}
\begin{split}
&|H\rangle\leftrightarrow\frac{1}{\sqrt{2}}(-|H\rangle+|V\rangle), \;\;
 |V\rangle\leftrightarrow\frac{1}{\sqrt{2}}(|H\rangle+|V\rangle).
\end{split}
\end{eqnarray}
The operations described above change $|\psi_0\rangle$ to
\begin{eqnarray}              \label{eq9}
\begin{split}
|\psi_1\rangle=&\frac{1}{\sqrt{2}}[
 \alpha_1|H_{1'}\rangle  (|H_{4}\rangle+|V_{4}\rangle)
+\alpha_2|H_{1'}\rangle  (|H_{3}\rangle\\&+|V_{3}\rangle)
+\alpha_3|V_{1} \rangle  (|H_{4}\rangle+|V_{4}\rangle)
+\alpha_4|V_{1} \rangle  (|H_{3}\rangle\\&+|V_{3}\rangle)
+\alpha_5|H_{2} \rangle  (|H_{4}\rangle+|V_{4}\rangle)
+\alpha_6|H_{2} \rangle  (|H_{3}\rangle\\&+|V_{3}\rangle)].
\end{split}
\end{eqnarray}

The photons in modes 2 and 3 ($1'$, 1, and 4) are subsequently combined at PBS$_3$ (PBS$_4$), and go through  HWP$^{67.5^\circ}$ (HWP$^{22.5^\circ}$). These operations transform $|\psi_1\rangle$ into
\begin{eqnarray}              \label{eq10}
\begin{split}
|\psi_2\rangle=&\frac{1}{2\sqrt{2}}[
 \alpha_1(|H_{8'}\rangle+|V_{8'}\rangle)(|H_{7}\rangle+|V_{7}\rangle+|H_{8}\rangle\\&-|V_{8}\rangle)
+\alpha_2(|H_{8'}\rangle+|V_{8'}\rangle)(-|H_{5}\rangle+|V_{5}\rangle\\&+|H_{6}\rangle+|V_{6}\rangle)
+\alpha_3(|H_{7}\rangle-|V_{7}\rangle)  (|H_{7}\rangle+|V_{7}\rangle\\&+|H_{8}\rangle-|V_{8}\rangle)
+\alpha_4(|H_{7}\rangle-|V_{7}\rangle)  (-|H_{5}\rangle\\&+|V_{5}\rangle+|H_{6}\rangle+|V_{6}\rangle)
+\alpha_5(-|H_{6}\rangle+|V_{6}\rangle)\\&\otimes (|H_{7}\rangle+|V_{7}\rangle+|H_{8}\rangle-|V_{8}\rangle)
+\alpha_6(-|H_{6}\rangle\\&+|V_{6}\rangle) (|V_{5}\rangle-|H_{5}\rangle+|H_{6}\rangle+|V_{6}\rangle)].
\end{split}
\end{eqnarray}

Finally, the photons emitted from modes 5 (6) and 8 (7) are mixed at PBS$_5$ (PBS$_6$). In addition, the photons emitted from mode $8'$ are split into $9'$ and $10'$ by PBS$_5$. After the photons in modes 10, $10'$, and 12 pass through HWP$^{45^\circ}$, the state of the system becomes
\begin{eqnarray}              \label{eq11}
\begin{split}
|\psi_3\rangle=&\frac{1}{2\sqrt{2}}[
 \alpha_1(|H_{9'}\rangle+|H_{10'}\rangle)
          (|H_{11}\rangle+|H_{12}\rangle\\&+|H_{9}\rangle-|H_{10}\rangle)
+\alpha_2 (|H_{9'}\rangle+|H_{10'}\rangle)
          (|V_{9}\rangle\\&-|V_{10}\rangle+|V_{12}\rangle+|V_{11}\rangle)
+\alpha_3 (|H_{11}\rangle\\&-|H_{12}\rangle)
          (|H_{11}\rangle+|H_{12}\rangle+|H_{9}\rangle-|H_{10}\rangle)\\&
+\alpha_4 (|H_{11}\rangle-|H_{12}\rangle)
          (|V_{9}\rangle-|V_{10}\rangle+|V_{12}\rangle\\&+|V_{11}\rangle)
+\alpha_5 (|V_{11}\rangle-|V_{12}\rangle)
          (|H_{11}\rangle+|H_{12}\rangle\\&+|H_{9}\rangle-|H_{10}\rangle)
+\alpha_6 (|V_{11}\rangle-|V_{12}\rangle)
          (|V_{9}\rangle\\&-|V_{10}\rangle+|V_{12}\rangle+|V_{11}\rangle)].
\end{split}
\end{eqnarray}

\begin{table*}[htb]
\centering \caption{Expected output values calculated for computational-logical-basis inputs.}
\begin{tabular}{ccccccc}
\hline  \hline
\multirow{4}{*}{Input}

& $\langle n_{H_{9'}}n_{H_{11}}\rangle$ & $\langle n_{H_{9'}}n_{V_{11}}\rangle$ & $\langle n_{H_{9}}n_{H_{11}}\rangle$ & $\langle n_{V_{9}}n_{H_{11}}\rangle$ & $\langle n_{H_{9}}n_{V_{11}}\rangle$ & $\langle n_{V_{9}}n_{V_{11}}\rangle$  \\


& $\langle n_{H_{10'}}n_{H_{12}}\rangle$ & $\langle n_{H_{10'}}n_{V_{12}}\rangle$ & $\langle n_{H_{10}}n_{H_{12}}\rangle$ & $\langle n_{V_{10}}n_{H_{12}}\rangle$ & $\langle n_{H_{10}}n_{V_{12}}\rangle$ & $\langle n_{V_{10}}n_{V_{12}}\rangle$  \\


& $\langle n_{H_{9'}}n_{H_{12}}\rangle$ & $\langle n_{H_{9'}}n_{V_{12}}\rangle$ & $-\langle n_{H_{9}}n_{H_{12}}\rangle$ & $-\langle n_{V_{9}}n_{H_{12}}\rangle$ & $-\langle n_{H_{9}}n_{V_{12}}\rangle$ & $-\langle n_{V_{9}}n_{V_{12}}\rangle$  \\


& $\langle n_{H_{10'}}n_{H_{11}}\rangle$ & $\langle n_{H_{10'}}n_{V_{11}}\rangle$ & $-\langle n_{H_{10}}n_{H_{11}}\rangle$ & $-\langle n_{V_{10}}n_{H_{11}}\rangle$ & $-\langle n_{H_{10}}n_{V_{11}}\rangle$ & $-\langle n_{V_{10}}n_{V_{11}}\rangle$  \\

\hline \hline
$|H_u\rangle|H\rangle$   & 1/8  &  0  &  0    & 0  &  0    & 0   \\

$|H_u\rangle|V\rangle$   &  0   &   1/8  & 0  & 0  &  0    & 0   \\

$|H_d\rangle|H\rangle$   & 0    &  0  &  1/8  & 0  &  0    & 0  \\

$|H_d\rangle|V\rangle$   & 0    &  0  &   0    & 1/8 &  0    & 0 \\

$|V_d\rangle|H\rangle$   & 0    &  0  &  0  & 0   &  1/8    & 0 \\

$|V_d\rangle|V\rangle$   & 0    &  0  &  0    &0   &  0    & 1/8 \\

\hline  \hline
\end{tabular}\label{table1}
\end{table*}

Table \ref{table1} shows that the quantum circuit presented in Fig. \ref{post-selected-2} completes a postselected partial-swap gate with a success probability of 1/4. The success probability of the gate can be  enhanced to 1/2 by applying a  phase shifter $P_\pi$ to mode 9 (10) when the photons are in modes 9 and 12 (10 and 11).  Here, the feedforward operation $P_\pi$ induces $|H_{9}\rangle\leftrightarrow-|H_{9}\rangle$ and $|V_{9}\rangle\leftrightarrow-|V_{9}\rangle$ ($|H_{10}\rangle\leftrightarrow-|H_{10}\rangle$ and $|V_{10}\rangle\leftrightarrow-|V_{10}\rangle$).

In the case of the construction of the three-photon Fredkin gate (see Fig. \ref{post-selected-1}), one finds that the leftmost PBS changes a system composed of photons $c$, $t_1$, and $t_2$ from the normalized arbitrary initial state $|\varphi_0\rangle$ into $|\varphi_1\rangle$. Here,
\begin{eqnarray}              \label{eq12}
\begin{split}
|\varphi_0\rangle=&\alpha_1|H\rangle_c|H\rangle_{t_1}|H\rangle_{t_2}+\alpha_2|H\rangle_c|H\rangle_{t_1}|V\rangle_{t_2}\\
                 &+\alpha_3|H\rangle_c|V\rangle_{t_1}|H\rangle_{t_2}+\alpha_4|H\rangle_c|V\rangle_{t_1}|V\rangle_{t_2}\\
                 &+\alpha_5|V\rangle_c|H\rangle_{t_1}|H\rangle_{t_2}+\alpha_6|V\rangle_c|H\rangle_{t_1}|V\rangle_{t_2}\\
                 &+\alpha_7|V\rangle_c|V\rangle_{t_1}|H\rangle_{t_2}+\alpha_8|V\rangle_c|V\rangle_{t_1}|V\rangle_{t_2},
\end{split}
\end{eqnarray}
\begin{eqnarray}              \label{eq13}
\begin{split}
|\varphi_1\rangle=&\alpha_1|H_{u}\rangle_c|H\rangle_{t_1}|H\rangle_{t_2}+\alpha_2|H_{u}\rangle_c|H\rangle_{t_1}|V\rangle_{t_2}\\
                 &+\alpha_3|H_{u}\rangle_c|V\rangle_{t_1}|H\rangle_{t_2}+\alpha_4|H_{u}\rangle_c|V\rangle_{t_1}|V\rangle_{t_2}\\
                 &+\alpha_5|V_{d}\rangle_c|H\rangle_{t_1}|H\rangle_{t_2}+\alpha_6|V_{d}\rangle_c|H\rangle_{t_1}|V\rangle_{t_2}\\
                 &+\alpha_7|V_{d}\rangle_c|V\rangle_{t_1}|H\rangle_{t_2}+\alpha_8|V_{d}\rangle_c|V\rangle_{t_1}|V\rangle_{t_2}.
\end{split}
\end{eqnarray}

After the three partial-swap gates (depicted in Fig. \ref{post-selected-2} without feedforwards) are applied, we obtain the 16 desired outcomes
$|\varphi_{9,11,11}^+\rangle$, $|\varphi_{9,11,11}^+\rangle$,
$|\varphi_{9,11,12}^-\rangle$, $|\varphi_{9,11,12}^-\rangle$,
$|\varphi_{9,12,11}^+\rangle$, $|\varphi_{9,12,11}^+\rangle$,
$|\varphi_{9,12,12}^-\rangle$, $|\varphi_{9,12,12}^-\rangle$,
$|\varphi_{10,11,11}^-\rangle$, $|\varphi_{10,11,11}^-\rangle$,
$|\varphi_{10,11,12}^+\rangle$, $|\varphi_{10,11,12}^+\rangle$,
$|\varphi_{10,12,11}^-\rangle$, $|\varphi_{10,12,11}^-\rangle$,
$|\varphi_{10,12,12}^+\rangle$, and $|\varphi_{10,12,12}^+\rangle$.
Here $|\varphi_{i,j,k}^{\pm}\rangle$, with $i\in\{9,10\}$ and $j,k\in\{11,12\}$, is given by
\begin{eqnarray}              \label{eq14}
\begin{split}
|\varphi_{i,j,k}^{\pm}\rangle=&\frac{1}{16\sqrt{2}}(
   \alpha_1|H_{i'}\rangle_c|H_{j}\rangle_{t_1}|H_{k}\rangle_{t_2}\\&
  +\alpha_2|H_{i'}\rangle_c|H_{j}\rangle_{t_1}|V_{k}\rangle_{t_2}\\&
  +\alpha_3|H_{i'}\rangle_c|V_{j}\rangle_{t_1}|H_{k}\rangle_{t_2}\\&
  +\alpha_4|H_{i'}\rangle_c|V_{j}\rangle_{t_1}|V_{k}\rangle_{t_2}\\&
\pm\alpha_5|V_{i}\rangle_c|H_{j}\rangle_{t_1}|H_{k}\rangle_{t_2}\\&
\pm\alpha_6|V_{i}\rangle_c|V_{j}\rangle_{t_1}|H_{k}\rangle_{t_2}\\&
\pm\alpha_7|V_{i}\rangle_c|H_{j}\rangle_{t_1}|V_{k}\rangle_{t_2}\\&
\pm\alpha_8|V_{i}\rangle_c|V_{j}\rangle_{t_1}|V_{k}\rangle_{t_2}).
\end{split}
\end{eqnarray}

Next, as shown in Fig. \ref{post-selected-1}, the rightmost PBS converges the photons into one mode, yielding
$|\tilde{\varphi}_{11,11,11}^+\rangle$, $|\tilde{\varphi}_{11,11,11}^+\rangle$,
$|\tilde{\varphi}_{11,11,12}^-\rangle$, $|\tilde{\varphi}_{11,11,12}^-\rangle$,
$|\tilde{\varphi}_{11,12,11}^+\rangle$, $|\tilde{\varphi}_{11,12,11}^+\rangle$,
$|\tilde{\varphi}_{11,12,12}^-\rangle$, $|\tilde{\varphi}_{11,12,12}^-\rangle$,
$|\tilde{\varphi}_{12,11,11}^-\rangle$, $|\tilde{\varphi}_{12,11,11}^-\rangle$,
$|\tilde{\varphi}_{12,11,12}^+\rangle$, $|\tilde{\varphi}_{12,11,12}^+\rangle$,
$|\tilde{\varphi}_{12,12,11}^-\rangle$, $|\tilde{\varphi}_{12,12,11}^-\rangle$,
$|\tilde{\varphi}_{12,12,12}^+\rangle$, and $|\tilde{\varphi}_{12,12,12}^+\rangle$.
Here, $|\tilde{\varphi}_{l,m,n}^{\pm}\rangle$ with  $l,m,n\in\{11,12\}$ is given by
\begin{eqnarray}              \label{eq15}
\begin{split}
|\tilde{\varphi}_{l,m,n}^{\pm}\rangle=&\frac{1}{16\sqrt{2}}(
   \alpha_1|H_{l}\rangle_c|H_{m}\rangle_{t_1}|H_{n}\rangle_{t_2}\\&
  +\alpha_2|H_{l}\rangle_c|H_{m}\rangle_{t_1}|V_{n}\rangle_{t_2}\\&
  +\alpha_3|H_{l}\rangle_c|V_{m}\rangle_{t_1}|H_{n}\rangle_{t_2}\\&
  +\alpha_4|H_{l}\rangle_c|V_{m}\rangle_{t_1}|V_{n}\rangle_{t_2}\\&
\pm\alpha_5|V_{l}\rangle_c|H_{m}\rangle_{t_1}|H_{n}\rangle_{t_2}\\&
\pm\alpha_6|V_{l}\rangle_c|V_{m}\rangle_{t_1}|H_{n}\rangle_{t_2}\\&
\pm\alpha_7|V_{l}\rangle_c|H_{m}\rangle_{t_1}|V_{n}\rangle_{t_2}\\&
\pm\alpha_8|V_{l}\rangle_c|V_{m}\rangle_{t_1}|V_{n}\rangle_{t_2}).
\end{split}
\end{eqnarray}


Finally, the minus signs in Eq. (\ref{eq15}) can be corrected by applying an HWP$^{0^\circ}$  to the mode  11 (12) of photon $c$ if the outgoing photons $c$, $t_1$, and $t_2$ are in modes 11, 11, and 12 or 11, 12, and 12 (12, 11, and 11 or 12, 12, and 11), respectively. Here, HWP$^{0^\circ}$ completes $|H\rangle\leftrightarrow|H\rangle$ and $|V\rangle\leftrightarrow-|V\rangle$.

Therefore, the quantum circuit presented in Fig. \ref{post-selected-1} completes a three-qubit Fredkin gate with a success probability of 1/32 (see Appendix for details), which beats the value of 1/192 given in Ref. \cite{Gong}. The number of optical components required for our scheme (a total of 50 optical components) is similar to that for the five-CNOT-based scheme (42 optical components) \cite{T-NatPhy,Liu}. In addition, five additional auxiliary entangled photon pairs are required for implementing a five-CNOT-based Fredkin gate \cite{Integrated-optics2,T-NatPhy,Liu}.

\begin{figure}[htpb]
\begin{center}
\includegraphics[width=8.6 cm,angle=0]{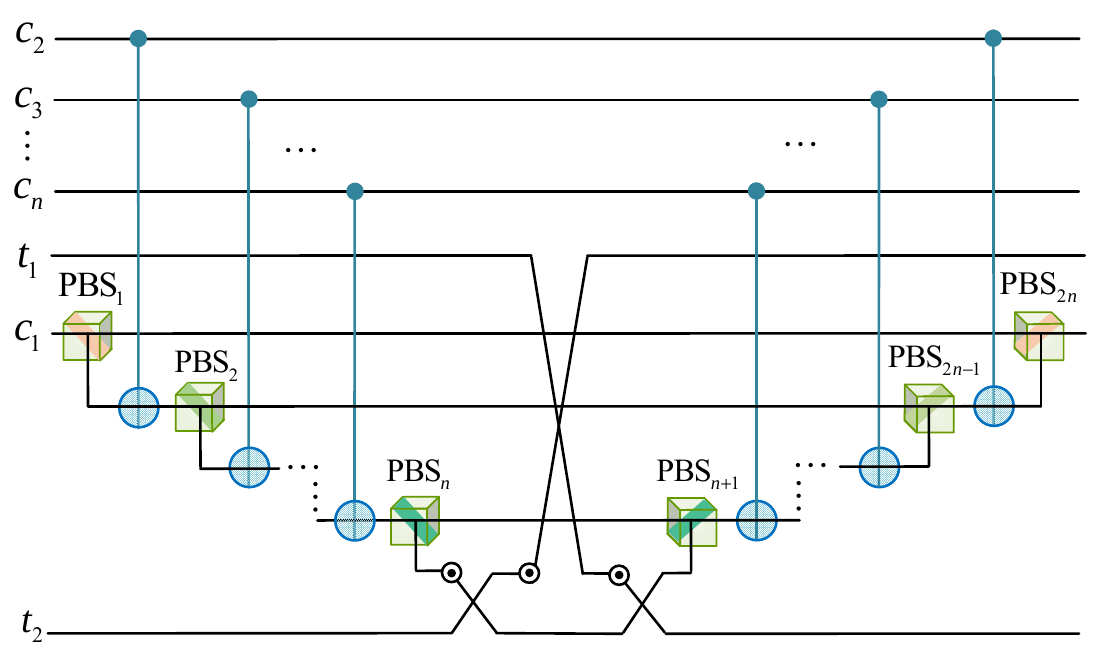}
\caption{(Color online) Schematic illustration of the optical implementation of an $n$-controlled-qubit Fredkin gate. } \label{n-implementation}
\end{center}
\end{figure}

The optical implementation of an $n$-controlled-qubit Fredkin gate with a success probability of $1/2^{4n+1}$ is depicted in Fig. \ref{n-implementation}. The $2n$ PBSs play the roles of $2n$ single-qudit gates to provide the extra spatial modes. CNOT gates can be optically realized with a  success probability of 1/4, assisted by entangled photon pairs \cite{Integrated-optics2,opti-CNOT}.

\section{Conclusion} \label{sec4}


In summary, we propose an alternative optimization of Fredkin gates using auxiliary states and partial-swap operations. Our scheme decreases the cost of a one-controlled-qubit Fredkin gate from the theoretical lower bound of five two-qubit gates \cite{Fredkin5-2} to three partial-swap gates using higher-dimensional Hilbert spaces. By extending the scheme to an $(n+2)$-dimensional subspace for the first control qubit, we show that three partial-swap gates and $2(n-1)$ CNOT gates supplemented with $2n$ single-qudit gates are sufficient to implement an $n$-controlled-qubit Fredkin gate. This gate circuit improves on previous results \cite{Barenco,T-NatPhy}. The construction procedure can further optimize a universal quantum circuit with higher-level systems and bridge the gap towards achieving the lower bound of $(4^n-3n-1)/4$ for any $n$-qubit quantum gate \cite{Shende}. The properties of the extra ancilla states need to be optimized for realizing quantum computing. For example, the computational qubits can be encoded as photonic polarizations (which provide low decoherence), and the extra ancilla qubits can  be encoded as spatial modes of a single photon (which are robust against bit-flip errors), or artificial atoms in a cavity (which provide a long coherence time) can be used as extra ancilla states. Alternatively, the computational qubits can  be encoded as electron-spin states of a nitrogen-vacancy center (with a coherence time of the order of milliseconds), and the extra ancilla qubits can be encoded as nuclear-spin states (with a coherence time of the order of seconds). Quantum computation using catalysis with higher-dimensional Hilbert spaces has been experimentally demonstrated in a linear optics system \cite{T-NatPhy} and a superconducting circuit \cite{F-superconduct}.

We also investigate linear optical implementations of Fredkin gates with the aid of a spatial degree of freedom. Without resorting to additional ancillary photons, we design a partial-swap gate with a success probability of 1/2, which beats the CNOT gate, which has a success probability of $1/9$ \cite{CNOT-BS1,CNOT-BS2,CNOT-BS3,CNOT-BS4}. Moreover, an optical Fredkin gate with an overall success probability of $1/32$ is implemented using the proposed three partial-swap gates. The latter construction improves on previous proposals \cite{Gong,construction1,construction2} in terms of the quantum resource cost and the probability of success of the gate.

Our optical implementation may be useful for  universal quantum computing with linear optics. The dominate imperfections for linear optical quantum computing are photon loss, detector inefficiency, and phase errors \cite{KLM,LOQC}. These optical errors can be reduced below the fault-tolerance threshold by measurement and error correction, which then allows the possibility of scalable optical quantum computing  \cite{LOQC,Loss-tolerant}.

\section*{Acknowledgments}

This study was supported by the National Natural Science Foundation of China under Grant No. 11604012, the Fundamental Research Funds for the Central Universities under Grants No. FRF-BR-17-004B and No. 230201506500024, and a grant from the China Scholarship Council.  L.-C.K. is supported by the Ministry of Education and the National Research Foundation Singapore.


\appendix*

\section{Implementation of linear optical Fredkin gate}

Following Sec. IIA, we encode the computational qubit with a polarization of the single photon in mode, i.e., $|0\rangle\equiv|H\rangle_d$ (black in Fig. \ref{post-selected-1}), $|1\rangle\equiv|V\rangle_d$ (also black in Fig. \ref{post-selected-1}). The third level (the additional state), $|2\rangle$, is encoded with the $H$-polarized component in   ``a new mode $u$'', i.e., $|2\rangle\equiv|H\rangle_u$ (brown in Fig. \ref{post-selected-1}), and this trick is completed by a PBS.

Based on Eqs. (\ref{eq6})-(\ref{eq11}) and Tab. \ref{table1}, one finds that the scheme in Fig. \ref{post-selected-2} completes the transformations
\begin{eqnarray}              \label{A1}
\begin{split}
&|H_u\rangle|H\rangle \rightarrow |H_{9'}\rangle|H_{11}\rangle, \;\;\,    |H_u\rangle|V\rangle \rightarrow |H_{9'}\rangle|V_{11}\rangle,\\
&|H_d\rangle|H\rangle \rightarrow |H_{9}\rangle|H_{11}\rangle,  \quad     |H_d\rangle|V\rangle \rightarrow |V_{9}\rangle|H_{11}\rangle, \\
&|V_d\rangle|H\rangle \rightarrow |H_{9}\rangle|V_{11}\rangle,  \quad\;\; |V_d\rangle|V\rangle \rightarrow |V_{9}\rangle|V_{11}\rangle,
\end{split}
\end{eqnarray}
or
\begin{eqnarray}              \label{A2}
\begin{split}
&|H_u\rangle|H\rangle \rightarrow |H_{10'}\rangle|H_{12}\rangle, \;\;\,    |H_u\rangle|V\rangle \rightarrow |H_{10'}\rangle|V_{12}\rangle,\\
&|H_d\rangle|H\rangle \rightarrow |H_{10}\rangle|H_{12}\rangle,  \quad     |H_d\rangle|V\rangle \rightarrow |V_{10}\rangle|H_{12}\rangle, \\
&|V_d\rangle|H\rangle \rightarrow |H_{10}\rangle|V_{12}\rangle,  \quad\;\; |V_d\rangle|V\rangle \rightarrow |V_{10}\rangle|V_{12}\rangle,
\end{split}
\end{eqnarray}
or
\begin{eqnarray}              \label{A3}
\begin{split}
&|H_u\rangle|H\rangle \rightarrow |H_{9'}\rangle|H_{12}\rangle, \;\;\;  |H_u\rangle|V\rangle \rightarrow |H_{9'}\rangle|V_{12}\rangle,\\
&|H_d\rangle|H\rangle \rightarrow -|H_{9}\rangle|H_{12}\rangle,  \;     |H_d\rangle|V\rangle \rightarrow -|V_{9}\rangle|H_{12}\rangle, \\
&|V_d\rangle|H\rangle \rightarrow -|H_{9}\rangle|V_{12}\rangle,  \;\;\; |V_d\rangle|V\rangle \rightarrow -|V_{9}\rangle|V_{12}\rangle,
\end{split}
\end{eqnarray}
or
\begin{eqnarray}              \label{A4}
\begin{split}
&|H_u\rangle|H\rangle \rightarrow |H_{10'}\rangle|H_{11}\rangle,\;\; |H_u\rangle|V\rangle \rightarrow |H_{10'}\rangle|V_{11}\rangle,\\
&|H_d\rangle|H\rangle \rightarrow -|H_{10}\rangle|H_{11}\rangle,     |H_d\rangle|V\rangle \rightarrow -|V_{10}\rangle|H_{11}\rangle, \\
&|V_d\rangle|H\rangle \rightarrow -|H_{10}\rangle|V_{11}\rangle,\;\; |V_d\rangle|V\rangle \rightarrow -|V_{10}\rangle|V_{11}\rangle.
\end{split}
\end{eqnarray}
Eqs. (\ref{A1}) and (\ref{A2}) realize a partial-swap gate, and Eqs. (\ref{A3}) and (\ref{A4}) can also realize a partial-swap gate if a  phase shifter $P_\pi$ is applied to mode 9 (10) when the photons are in modes 9 and 12 (10 and 11).

Next, we demonstrate the three-photon Fredkin gate presented in Fig. \ref{post-selected-1} based on three partial-swap gates.
As shown in Fig. \ref{post-selected-1}, the leftmost PBS transforms the system from an arbitrary normalized initial state $|\varphi_0\rangle$ into $|\varphi_1\rangle$. Here,
\begin{eqnarray}              \label{A5}
\begin{split}
|\varphi_0\rangle=&\alpha_1|H\rangle_c|H\rangle_{t_1}|H\rangle_{t_2}+\alpha_2|H\rangle_c|H\rangle_{t_1}|V\rangle_{t_2}\\
                 &+\alpha_3|H\rangle_c|V\rangle_{t_1}|H\rangle_{t_2}+\alpha_4|H\rangle_c|V\rangle_{t_1}|V\rangle_{t_2}\\
                 &+\alpha_5|V\rangle_c|H\rangle_{t_1}|H\rangle_{t_2}+\alpha_6|V\rangle_c|H\rangle_{t_1}|V\rangle_{t_2}\\
                 &+\alpha_7|V\rangle_c|V\rangle_{t_1}|H\rangle_{t_2}+\alpha_8|V\rangle_c|V\rangle_{t_1}|V\rangle_{t_2},
\end{split}
\end{eqnarray}
\begin{eqnarray}              \label{A6}
\begin{split}
|\varphi_1\rangle=&\alpha_1|H_{u}\rangle_c|H\rangle_{t_1}|H\rangle_{t_2}+\alpha_2|H_{u}\rangle_c|H\rangle_{t_1}|V\rangle_{t_2}\\
                 &+\alpha_3|H_{u}\rangle_c|V\rangle_{t_1}|H\rangle_{t_2}+\alpha_4|H_{u}\rangle_c|V\rangle_{t_1}|V\rangle_{t_2}\\
                 &+\alpha_5|V_{d}\rangle_c|H\rangle_{t_1}|H\rangle_{t_2}+\alpha_6|V_{d}\rangle_c|H\rangle_{t_1}|V\rangle_{t_2}\\
                 &+\alpha_7|V_{d}\rangle_c|V\rangle_{t_1}|H\rangle_{t_2}+\alpha_8|V_{d}\rangle_c|V\rangle_{t_1}|V\rangle_{t_2}.
\end{split}
\end{eqnarray}
where $H_u$ ($V_d$) represents an $H$- ($V$-) polarized photon emitted from the mode $u$ ($d$).

After photons $c$ and $t_2$ pass through the first partial-swap gate, we obtain two desired states,
\begin{eqnarray}              \label{A7}
\begin{split}
|\varphi_{2a}\rangle=&\frac{1}{2\sqrt{2}}(\alpha_1|H_{9'}\rangle_{c}|H\rangle_{t_1}|H_{11}\rangle_{t_2}\\&+\alpha_2|H_{9'}\rangle_{c}|H\rangle_{t_1}|V_{11}\rangle_{t_2}
                  \\& +\alpha_3|H_{9'}\rangle_{c}|V\rangle_{t_1}|H_{11}\rangle_{t_2}\\&+\alpha_4|H_{9'}\rangle_{c}|V\rangle_{t_1}|V_{11}\rangle_{t_2}
                  \\& +\alpha_5|H_{9}\rangle_{c}|H\rangle_{t_1}|V_{11}\rangle_{t_2}\\&+\alpha_6|V_{9}\rangle_{c}|H\rangle_{t_1}|V_{11}\rangle_{t_2}
                  \\& +\alpha_7|H_{9}\rangle_{c}|V\rangle_{t_1}|V_{11}\rangle_{t_2}\\&+\alpha_8|V_{9}\rangle_{c}|V\rangle_{t_1}|V_{11}\rangle_{t_2}),
\end{split}
\end{eqnarray}
\begin{eqnarray}              \label{A8}
\begin{split}
|\varphi_{2b}\rangle=&\frac{1}{2\sqrt{2}}(\alpha_1|H_{10'}\rangle_{c}|H\rangle_{t_1}|H_{12}\rangle_{t_2}\\&+\alpha_2|H_{10'}\rangle_{c}|H\rangle_{t_1}|V_{12}\rangle_{t_2}
                   \\&+\alpha_3|H_{10'}\rangle_{c}|V\rangle_{t_1}|H_{12}\rangle_{t_2}\\&+\alpha_4|H_{10'}\rangle_{c}|V\rangle_{t_1}|V_{12}\rangle_{t_2}
                   \\&+\alpha_5|H_{10}\rangle_{c}|H\rangle_{t_1}|V_{12}\rangle_{t_2}\\&+\alpha_6|V_{10}\rangle_{c}|H\rangle_{t_1}|V_{12}\rangle_{t_2}
                   \\&+\alpha_7|H_{10}\rangle_{c}|V\rangle_{t_1}|V_{12}\rangle_{t_2}\\&+\alpha_8|V_{10}\rangle_{c}|V\rangle_{t_1}|V_{12}\rangle_{t_2}).
\end{split}
\end{eqnarray}

The second partial-swap gate, acting on $c$ and $t_1$, yields four desired states,
\begin{eqnarray}              \label{A9}
\begin{split}
|\varphi_{3aa}\rangle=&\frac{1}{8}(\alpha_1|H_{9'}\rangle_{c}|H_{11}\rangle_{t_1}|H_{11}\rangle_{t_2}\\&+\alpha_2|H_{9'}\rangle_{c}|H_{11}\rangle_{t_1}|V_{11}\rangle_{t_2}
                  \\&+\alpha_3|H_{9'}\rangle_{c}|V_{11}\rangle_{t_1}|H_{11}\rangle_{t_2}\\&+\alpha_4|H_{9'}\rangle_{c}|V_{11}\rangle_{t_1}|V_{11}\rangle_{t_2}
                  \\&+\alpha_5|H_{9}\rangle_{c}|H_{11}\rangle_{t_1}|V_{11}\rangle_{t_2}\\&+\alpha_6|H_{9}\rangle_{c}|V_{11}\rangle_{t_1}|V_{11}\rangle_{t_2}
                  \\&+\alpha_7|V_{9}\rangle_{c}|H_{11}\rangle_{t_1}|V_{11}\rangle_{t_2}\\&+\alpha_8|V_{9}\rangle_{c}|V_{11}\rangle_{t_1}|V_{11}\rangle_{t_2}),
\end{split}
\end{eqnarray}

\begin{eqnarray}              \label{A10}
\begin{split}
|\varphi_{3ab}\rangle=&\frac{1}{8}(\alpha_1|H_{10'}\rangle_{c}|H_{12}\rangle_{t_1}|H_{11}\rangle_{t_2}\\&+\alpha_2|H_{10'}\rangle_{c}|H_{12}\rangle_{t_1}|V_{11}\rangle_{t_2}
                  \\&+\alpha_3|H_{10'}\rangle_{c}|V_{12}\rangle_{t_1}|H_{11}\rangle_{t_2}\\&+\alpha_4|H_{10'}\rangle_{c}|V_{12}\rangle_{t_1}|V_{11}\rangle_{t_2}
                  \\&+\alpha_5|H_{10}\rangle_{c}|H_{12}\rangle_{t_1}|V_{11}\rangle_{t_2}\\&+\alpha_6|H_{10}\rangle_{c}|V_{12}\rangle_{t_1}|V_{11}\rangle_{t_2}
                  \\&+\alpha_7|V_{10}\rangle_{c}|H_{12}\rangle_{t_1}|V_{11}\rangle_{t_2}\\&+\alpha_8|V_{10}\rangle_{c}|V_{12}\rangle_{t_1}|V_{11}\rangle_{t_2}),
\end{split}
\end{eqnarray}

\begin{eqnarray}              \label{A11}
\begin{split}
|\varphi_{3ba}\rangle=&\frac{1}{8}(\alpha_1|H_{9'}\rangle_{c}|H_{11}\rangle_{t_1}|H_{12}\rangle_{t_2}\\&+\alpha_2|H_{9'}\rangle_{c}|H_{11}\rangle_{t_1}|V_{12}\rangle_{t_2}
                  \\&+\alpha_3|H_{9'}\rangle_{c}|V_{11}\rangle_{t_1}|H_{12}\rangle_{t_2}\\&+\alpha_4|H_{9'}\rangle_{c}|V_{11}\rangle_{t_1}|V_{12}\rangle_{t_2}
                  \\&+\alpha_5|H_{9}\rangle_{c}|H_{11}\rangle_{t_1}|V_{12}\rangle_{t_2}\\&+\alpha_6|H_{9}\rangle_{c}|V_{11}\rangle_{t_1}|V_{12}\rangle_{t_2}
                  \\&+\alpha_7|V_{9}\rangle_{c}|H_{11}\rangle_{t_1}|V_{12}\rangle_{t_2}\\&+\alpha_8|V_{9}\rangle_{c}|V_{11}\rangle_{t_1}|V_{12}\rangle_{t_2}),
\end{split}
\end{eqnarray}
\begin{eqnarray}              \label{A12}
\begin{split}
|\varphi_{3bb}\rangle=&\frac{1}{8}(\alpha_1|H_{10'}\rangle_{c}|H_{12}\rangle_{t_1}|H_{12}\rangle_{t_2}\\&+\alpha_2|H_{10'}\rangle_{c}|H_{12}\rangle_{t_1}|V_{12}\rangle_{t_2}
                  \\&+\alpha_3|H_{10'}\rangle_{c}|V_{12}\rangle_{t_1}|H_{12}\rangle_{t_2}\\&+\alpha_4|H_{10'}\rangle_{c}|V_{12}\rangle_{t_1}|V_{12}\rangle_{t_2}
                 \\& +\alpha_5|H_{10}\rangle_{c}|H_{12}\rangle_{t_1}|V_{12}\rangle_{t_2}\\&+\alpha_6|H_{10}\rangle_{c}|V_{12}\rangle_{t_1}|V_{12}\rangle_{t_2}
                  \\&+\alpha_7|V_{10}\rangle_{c}|H_{12}\rangle_{t_1}|V_{12}\rangle_{t_2}\\&+\alpha_8|V_{10}\rangle_{c}|V_{12}\rangle_{t_1}|V_{12}\rangle_{t_2}).
\end{split}
\end{eqnarray}

The third partial-swap gate, acting on $c$ and $t_2$, yields 16 desired outcomes,
\begin{eqnarray}              \label{A13}
\begin{split}
|\varphi_{9,11,11}^{+}\rangle=&\frac{1}{16\sqrt{2}}(
   \alpha_1|H_{9'}\rangle_c|H_{11}\rangle_{t_1}|H_{11}\rangle_{t_2}
  \\&+\alpha_2|H_{9'}\rangle_c |H_{11}\rangle_{t_1}|V_{11}\rangle_{t_2}
  \\&+\alpha_3|H_{9'}\rangle_c|V_{11}\rangle_{t_1}|H_{11}\rangle_{t_2}
  \\&+\alpha_4|H_{9'}\rangle_c|V_{11}\rangle_{t_1}|V_{11}\rangle_{t_2}
  \\&+\alpha_5|V_{9}\rangle_c|H_{11}\rangle_{t_1}|H_{11}\rangle_{t_2}
  \\&+\alpha_6|V_{9}\rangle_c|V_{11}\rangle_{t_1}|H_{11}\rangle_{t_2}
  \\&+\alpha_7|V_{9}\rangle_c|H_{11}\rangle_{t_1}|V_{11}\rangle_{t_2}
  \\&+\alpha_8|V_{9}\rangle_c|V_{11}\rangle_{t_1}|V_{11}\rangle_{t_2}),
\end{split}
\end{eqnarray}
\begin{eqnarray}              \label{A14}
\begin{split}
|\varphi_{10,11,12}^{+}\rangle=&\frac{1}{16\sqrt{2}}(
   \alpha_1|H_{10'}\rangle_c|H_{11}\rangle_{t_1}|H_{12}\rangle_{t_2}
  \\&+\alpha_2|H_{10'}\rangle_c|H_{11}\rangle_{t_1}|V_{12}\rangle_{t_2}
  \\&+\alpha_3|H_{10'}\rangle_c|V_{11}\rangle_{t_1}|H_{12}\rangle_{t_2}
  \\&+\alpha_4|H_{10'}\rangle_c|V_{11}\rangle_{t_1}|V_{12}\rangle_{t_2}
  \\&+\alpha_5|V_{10}\rangle_c|H_{11}\rangle_{t_1}|H_{12}\rangle_{t_2}
  \\&+\alpha_6|V_{10}\rangle_c|V_{11}\rangle_{t_1}|H_{12}\rangle_{t_2}
  \\&+\alpha_7|V_{10}\rangle_c|H_{11}\rangle_{t_1}|V_{12}\rangle_{t_2}
  \\&+\alpha_8|V_{10}\rangle_c|V_{11}\rangle_{t_1}|V_{12}\rangle_{t_2}),
\end{split}
\end{eqnarray}
\begin{eqnarray}              \label{A15}
\begin{split}
|\varphi_{9,11,12}^{-}\rangle=&\frac{1}{16\sqrt{2}}(
   \alpha_1|H_{9'}\rangle_c|H_{11}\rangle_{t_1}|H_{12}\rangle_{t_2}
  \\&+\alpha_2|H_{9'}\rangle_c|H_{11}\rangle_{t_1}|V_{12}\rangle_{t_2}
  \\&+\alpha_3|H_{9'}\rangle_c|V_{11}\rangle_{t_1}|H_{12}\rangle_{t_2}
  \\&+\alpha_4|H_{9'}\rangle_c|V_{11}\rangle_{t_1}|V_{12}\rangle_{t_2}
  \\&-\alpha_5|V_{9}\rangle_c|H_{11}\rangle_{t_1}|H_{12}\rangle_{t_2}
  \\&-\alpha_6|V_{9}\rangle_c|V_{11}\rangle_{t_1}|H_{12}\rangle_{t_2}
  \\&-\alpha_7|V_{9}\rangle_c|H_{11}\rangle_{t_1}|V_{12}\rangle_{t_2}
  \\&-\alpha_8|V_{9}\rangle_c|V_{11}\rangle_{t_1}|V_{12}\rangle_{t_2}),
\end{split}
\end{eqnarray}
\begin{eqnarray}              \label{A16}
\begin{split}
|\varphi_{10,11,11}^{-}\rangle=&\frac{1}{16\sqrt{2}}(
   \alpha_1|H_{10'}\rangle_c|H_{11}\rangle_{t_1}|H_{11}\rangle_{t_2}
  \\&+\alpha_2|H_{10'}\rangle_c|H_{11}\rangle_{t_1}|V_{11}\rangle_{t_2}
  \\&+\alpha_3|H_{10'}\rangle_c|V_{11}\rangle_{t_1}|H_{11}\rangle_{t_2}
  \\&+\alpha_4|H_{10'}\rangle_c|V_{11}\rangle_{t_1}|V_{11}\rangle_{t_2}
  \\&-\alpha_5|V_{10}\rangle_c|H_{11}\rangle_{t_1}|H_{11}\rangle_{t_2}
  \\&-\alpha_6|V_{10}\rangle_c|V_{11}\rangle_{t_1}|H_{11}\rangle_{t_2}
  \\&-\alpha_7|V_{10}\rangle_c|H_{11}\rangle_{t_1}|V_{11}\rangle_{t_2}
  \\&-\alpha_8|V_{10}\rangle_c|V_{11}\rangle_{t_1}|V_{11}\rangle_{t_2}),
\end{split}
\end{eqnarray}
\begin{eqnarray}              \label{A17}
\begin{split}
|\varphi_{9,12,11}^{+}\rangle=&\frac{1}{16\sqrt{2}}(
   \alpha_1|H_{9'}\rangle_c|H_{12}\rangle_{t_1}|H_{11}\rangle_{t_2}
  \\&+\alpha_2|H_{9'}\rangle_c|H_{12}\rangle_{t_1}|V_{11}\rangle_{t_2}
  \\&+\alpha_3|H_{9'}\rangle_c|V_{12}\rangle_{t_1}|H_{11}\rangle_{t_2}
  \\&+\alpha_4|H_{9'}\rangle_c|V_{12}\rangle_{t_1}|V_{11}\rangle_{t_2}
  \\&+\alpha_5|V_{9}\rangle_c|H_{12}\rangle_{t_1}|H_{11}\rangle_{t_2}
  \\&+\alpha_6|V_{9}\rangle_c|V_{12}\rangle_{t_1}|H_{11}\rangle_{t_2}
  \\&+\alpha_7|V_{9}\rangle_c|H_{12}\rangle_{t_1}|V_{11}\rangle_{t_2}
  \\&+\alpha_8|V_{9}\rangle_c|V_{12}\rangle_{t_1}|V_{11}\rangle_{t_2}),
\end{split}
\end{eqnarray}
\begin{eqnarray}              \label{A18}
\begin{split}
|\varphi_{10,12,12}^{+}\rangle=&\frac{1}{16\sqrt{2}}(
   \alpha_1|H_{10'}\rangle_c|H_{12}\rangle_{t_1}|H_{12}\rangle_{t_2}
  \\&+\alpha_2|H_{10'}\rangle_c|H_{12}\rangle_{t_1}|V_{12}\rangle_{t_2}
  \\&+\alpha_3|H_{10'}\rangle_c|V_{12}\rangle_{t_1}|H_{12}\rangle_{t_2}
  \\&+\alpha_4|H_{10'}\rangle_c|V_{12}\rangle_{t_1}|V_{12}\rangle_{t_2}
  \\&+\alpha_5|V_{10}\rangle_c|H_{12}\rangle_{t_1}|H_{12}\rangle_{t_2}
  \\&+\alpha_6|V_{10}\rangle_c|V_{12}\rangle_{t_1}|H_{12}\rangle_{t_2}
  \\&+\alpha_7|V_{10}\rangle_c|H_{12}\rangle_{t_1}|V_{12}\rangle_{t_2}
  \\&+\alpha_8|V_{10}\rangle_c|V_{12}\rangle_{t_1}|V_{12}\rangle_{t_2}),
\end{split}
\end{eqnarray}
\begin{eqnarray}              \label{A19}
\begin{split}
|\varphi_{9,12,12}^{-}\rangle=&\frac{1}{16\sqrt{2}}(
   \alpha_1|H_{9'}\rangle_c|H_{12}\rangle_{t_1}|H_{12}\rangle_{t_2}
  \\&+\alpha_2|H_{9'}\rangle_c|H_{12}\rangle_{t_1}|V_{12}\rangle_{t_2}
  \\&+\alpha_3|H_{9'}\rangle_c|V_{12}\rangle_{t_1}|H_{12}\rangle_{t_2}
  \\&+\alpha_4|H_{9'}\rangle_c|V_{12}\rangle_{t_1}|V_{12}\rangle_{t_2}
  \\&-\alpha_5|V_{9}\rangle_c|H_{12}\rangle_{t_1}|H_{12}\rangle_{t_2}
  \\&-\alpha_6|V_{9}\rangle_c|V_{12}\rangle_{t_1}|H_{12}\rangle_{t_2}
  \\&-\alpha_7|V_{9}\rangle_c|H_{12}\rangle_{t_1}|V_{12}\rangle_{t_2}
  \\&-\alpha_8|V_{9}\rangle_c|V_{12}\rangle_{t_1}|V_{12}\rangle_{t_2}),
\end{split}
\end{eqnarray}
\begin{eqnarray}              \label{A20}
\begin{split}
|\varphi_{10,12,11}^{-}\rangle=&\frac{1}{16\sqrt{2}}(
   \alpha_1|H_{10'}\rangle_c|H_{12}\rangle_{t_1}|H_{11}\rangle_{t_2}
  \\&+\alpha_2|H_{10'}\rangle_c|H_{12}\rangle_{t_1}|V_{11}\rangle_{t_2}
  \\&+\alpha_3|H_{10'}\rangle_c|V_{12}\rangle_{t_1}|H_{11}\rangle_{t_2}
  \\&+\alpha_4|H_{10'}\rangle_c|V_{12}\rangle_{t_1}|V_{11}\rangle_{t_2}
  \\&-\alpha_5|V_{10}\rangle_c|H_{12}\rangle_{t_1}|H_{11}\rangle_{t_2}
  \\&-\alpha_6|V_{10}\rangle_c|V_{12}\rangle_{t_1}|H_{11}\rangle_{t_2}
  \\&-\alpha_7|V_{10}\rangle_c|H_{12}\rangle_{t_1}|V_{11}\rangle_{t_2}
  \\&-\alpha_8|V_{10}\rangle_c|V_{12}\rangle_{t_1}|V_{11}\rangle_{t_2}),
\end{split}
\end{eqnarray}
\begin{eqnarray}              \label{A21}
\begin{split}
|\varphi_{9,11,11}^{+}\rangle=&\frac{1}{16\sqrt{2}}(
   \alpha_1|H_{9'}\rangle_c|H_{11}\rangle_{t_1}|H_{11}\rangle_{t_2}
  \\&+\alpha_2|H_{9'}\rangle_c|H_{11}\rangle_{t_1}|V_{11}\rangle_{t_2}
  \\&+\alpha_3|H_{9'}\rangle_c|V_{11}\rangle_{t_1}|H_{11}\rangle_{t_2}
  \\&+\alpha_4|H_{9'}\rangle_c|V_{11}\rangle_{t_1}|V_{11}\rangle_{t_2}
  \\&+\alpha_5|V_{9}\rangle_c|H_{11}\rangle_{t_1}|H_{11}\rangle_{t_2}
  \\&+\alpha_6|V_{9}\rangle_c|V_{11}\rangle_{t_1}|H_{11}\rangle_{t_2}
  \\&+\alpha_7|V_{9}\rangle_c|H_{11}\rangle_{t_1}|V_{11}\rangle_{t_2}
  \\&+\alpha_8|V_{9}\rangle_c|V_{11}\rangle_{t_1}|V_{11}\rangle_{t_2}),
\end{split}
\end{eqnarray}
\begin{eqnarray}              \label{A22}
\begin{split}
|\varphi_{10,11,12}^{+}\rangle=&\frac{1}{16\sqrt{2}}(
   \alpha_1|H_{10'}\rangle_c|H_{11}\rangle_{t_1}|H_{12}\rangle_{t_2}
  \\&+\alpha_2|H_{10'}\rangle_c|H_{11}\rangle_{t_1}|V_{12}\rangle_{t_2}
  \\&+\alpha_3|H_{10'}\rangle_c|V_{11}\rangle_{t_1}|H_{12}\rangle_{t_2}
  \\&+\alpha_4|H_{10'}\rangle_c|V_{11}\rangle_{t_1}|V_{12}\rangle_{t_2}
  \\&+\alpha_5|V_{10}\rangle_c|H_{11}\rangle_{t_1}|H_{12}\rangle_{t_2}
  \\&+\alpha_6|V_{10}\rangle_c|V_{11}\rangle_{t_1}|H_{12}\rangle_{t_2}
  \\&+\alpha_7|V_{10}\rangle_c|H_{11}\rangle_{t_1}|V_{12}\rangle_{t_2}
  \\&+\alpha_8|V_{10}\rangle_c|V_{11}\rangle_{t_1}|V_{12}\rangle_{t_2}),
\end{split}
\end{eqnarray}
\begin{eqnarray}              \label{A23}
\begin{split}
|\varphi_{9,11,12}^{-}\rangle=&\frac{1}{16\sqrt{2}}(
   \alpha_1|H_{9'}\rangle_c|H_{11}\rangle_{t_1}|H_{12}\rangle_{t_2}
  \\&+\alpha_2|H_{9'}\rangle_c|H_{11}\rangle_{t_1}|V_{12}\rangle_{t_2}
  \\&+\alpha_3|H_{9'}\rangle_c|V_{11}\rangle_{t_1}|H_{12}\rangle_{t_2}
  \\&+\alpha_4|H_{9'}\rangle_c|V_{11}\rangle_{t_1}|V_{12}\rangle_{t_2}
  \\&-\alpha_5|V_{9}\rangle_c|H_{11}\rangle_{t_1}|H_{12}\rangle_{t_2}
  \\&-\alpha_6|V_{9}\rangle_c|V_{11}\rangle_{t_1}|H_{12}\rangle_{t_2}
  \\&-\alpha_7|V_{9}\rangle_c|H_{11}\rangle_{t_1}|V_{12}\rangle_{t_2}
  \\&-\alpha_8|V_{9}\rangle_c|V_{11}\rangle_{t_1}|V_{12}\rangle_{t_2}),
\end{split}
\end{eqnarray}
\begin{eqnarray}              \label{A24}
\begin{split}
|\varphi_{10,11,11}^{-}\rangle=&\frac{1}{16\sqrt{2}}(
   \alpha_1|H_{10'}\rangle_c|H_{11}\rangle_{t_1}|H_{11}\rangle_{t_2}
  \\&+\alpha_2|H_{10'}\rangle_c|H_{11}\rangle_{t_1}|V_{11}\rangle_{t_2}
  \\&+\alpha_3|H_{10'}\rangle_c|V_{11}\rangle_{t_1}|H_{11}\rangle_{t_2}
  \\&+\alpha_4|H_{10'}\rangle_c|V_{11}\rangle_{t_1}|V_{11}\rangle_{t_2}
  \\&-\alpha_5|V_{10}\rangle_c|H_{11}\rangle_{t_1}|H_{11}\rangle_{t_2}
  \\&-\alpha_6|V_{10}\rangle_c|V_{11}\rangle_{t_1}|H_{11}\rangle_{t_2}
  \\&-\alpha_7|V_{10}\rangle_c|H_{11}\rangle_{t_1}|V_{11}\rangle_{t_2}
  \\&-\alpha_8|V_{10}\rangle_c|V_{11}\rangle_{t_1}|V_{11}\rangle_{t_2}),
\end{split}
\end{eqnarray}
\begin{eqnarray}              \label{A25}
\begin{split}
|\varphi_{9,12,11}^{+}\rangle=&\frac{1}{16\sqrt{2}}(
   \alpha_1|H_{9'}\rangle_c|H_{12}\rangle_{t_1}|H_{11}\rangle_{t_2}
  \\&+\alpha_2|H_{9'}\rangle_c|H_{12}\rangle_{t_1}|V_{11}\rangle_{t_2}
  \\&+\alpha_3|H_{9'}\rangle_c|V_{12}\rangle_{t_1}|H_{11}\rangle_{t_2}
  \\&+\alpha_4|H_{9'}\rangle_c|V_{12}\rangle_{t_1}|V_{11}\rangle_{t_2}
  \\&+\alpha_5|V_{9}\rangle_c|H_{12}\rangle_{t_1}|H_{11}\rangle_{t_2}
  \\&+\alpha_6|V_{9}\rangle_c|V_{12}\rangle_{t_1}|H_{11}\rangle_{t_2}
  \\&+\alpha_7|V_{9}\rangle_c|H_{12}\rangle_{t_1}|V_{11}\rangle_{t_2}
  \\&+\alpha_8|V_{9}\rangle_c|V_{12}\rangle_{t_1}|V_{11}\rangle_{t_2}),
\end{split}
\end{eqnarray}
\begin{eqnarray}              \label{A26}
\begin{split}
|\varphi_{10,12,12}^{+}\rangle=&\frac{1}{16\sqrt{2}}(
   \alpha_1|H_{10'}\rangle_c|H_{12}\rangle_{t_1}|H_{12}\rangle_{t_2}
  \\&+\alpha_2|H_{10'}\rangle_c|H_{12}\rangle_{t_1}|V_{12}\rangle_{t_2}
  \\&+\alpha_3|H_{10'}\rangle_c|V_{12}\rangle_{t_1}|H_{12}\rangle_{t_2}
  \\&+\alpha_4|H_{10'}\rangle_c|V_{12}\rangle_{t_1}|V_{12}\rangle_{t_2}
  \\&+\alpha_5|V_{10}\rangle_c|H_{12}\rangle_{t_1}|H_{12}\rangle_{t_2}
  \\&+\alpha_6|V_{10}\rangle_c|V_{12}\rangle_{t_1}|H_{12}\rangle_{t_2}
  \\&+\alpha_7|V_{10}\rangle_c|H_{12}\rangle_{t_1}|V_{12}\rangle_{t_2}
  \\&+\alpha_8|V_{10}\rangle_c|V_{12}\rangle_{t_1}|V_{12}\rangle_{t_2}),
\end{split}
\end{eqnarray}
\begin{eqnarray}              \label{A27}
\begin{split}
|\varphi_{9,12,12}^{-}\rangle=&\frac{1}{16\sqrt{2}}(
   \alpha_1|H_{9'}\rangle_c|H_{12}\rangle_{t_1}|H_{12}\rangle_{t_2}
  \\&+\alpha_2|H_{9'}\rangle_c|H_{12}\rangle_{t_1}|V_{12}\rangle_{t_2}
  \\&+\alpha_3|H_{9'}\rangle_c|V_{12}\rangle_{t_1}|H_{12}\rangle_{t_2}
  \\&+\alpha_4|H_{9'}\rangle_c|V_{12}\rangle_{t_1}|V_{12}\rangle_{t_2}
  \\&-\alpha_5|V_{9}\rangle_c|H_{12}\rangle_{t_1}|H_{12}\rangle_{t_2}
  \\&-\alpha_6|V_{9}\rangle_c|V_{12}\rangle_{t_1}|H_{12}\rangle_{t_2}
  \\&-\alpha_7|V_{9}\rangle_c|H_{12}\rangle_{t_1}|V_{12}\rangle_{t_2}
  \\&-\alpha_8|V_{9}\rangle_c|V_{12}\rangle_{t_1}|V_{12}\rangle_{t_2}),
\end{split}
\end{eqnarray}
\begin{eqnarray}              \label{A28}
\begin{split}
|\varphi_{10,12,11}^{-}\rangle=&\frac{1}{16\sqrt{2}}(
   \alpha_1|H_{10'}\rangle_c|H_{12}\rangle_{t_1}|H_{11}\rangle_{t_2}
  \\&+\alpha_2|H_{10'}\rangle_c|H_{12}\rangle_{t_1}|V_{11}\rangle_{t_2}
  \\&+\alpha_3|H_{10'}\rangle_c|V_{12}\rangle_{t_1}|H_{11}\rangle_{t_2}
  \\&+\alpha_4|H_{10'}\rangle_c|V_{12}\rangle_{t_1}|V_{11}\rangle_{t_2}
  \\&-\alpha_5|V_{10}\rangle_c|H_{12}\rangle_{t_1}|H_{11}\rangle_{t_2}
  \\&-\alpha_6|V_{10}\rangle_c|V_{12}\rangle_{t_1}|H_{11}\rangle_{t_2}
  \\&-\alpha_7|V_{10}\rangle_c|H_{12}\rangle_{t_1}|V_{11}\rangle_{t_2}
  \\&-\alpha_8|V_{10}\rangle_c|V_{12}\rangle_{t_1}|V_{11}\rangle_{t_2}).
\end{split}
\end{eqnarray}

Subsequently, the rightmost PBS converges the photons into one controlled mode ($|H_{9'}\rangle_c\rightarrow |H_{11}\rangle_c, |V_{9}\rangle_c\rightarrow |V_{11}\rangle_c, |H_{10'}\rangle_c\rightarrow |H_{12}\rangle_c$ and $|V_{10}\rangle_c\rightarrow |V_{12}\rangle_c$), yielding
$|\tilde{\varphi}_{11,11,11}^+\rangle$, $|\tilde{\varphi}_{11,11,11}^+\rangle$,
$|\tilde{\varphi}_{11,11,12}^-\rangle$, $|\tilde{\varphi}_{11,11,12}^-\rangle$,
$|\tilde{\varphi}_{11,12,11}^+\rangle$, $|\tilde{\varphi}_{11,12,11}^+\rangle$,
$|\tilde{\varphi}_{11,12,12}^-\rangle$, $|\tilde{\varphi}_{11,12,12}^-\rangle$,
$|\tilde{\varphi}_{12,11,11}^-\rangle$, $|\tilde{\varphi}_{12,11,11}^-\rangle$,
$|\tilde{\varphi}_{12,11,12}^+\rangle$, $|\tilde{\varphi}_{12,11,12}^+\rangle$,
$|\tilde{\varphi}_{12,12,11}^-\rangle$, $|\tilde{\varphi}_{12,12,11}^-\rangle$,
$|\tilde{\varphi}_{12,12,12}^+\rangle$ and $|\tilde{\varphi}_{12,12,12}^+\rangle$.
Here $|\tilde{\varphi}_{l,m,n}^{\pm}\rangle$, with  $l,m,n\in\{11,12\}$, is given by
\begin{eqnarray}              \label{A29}
\begin{split}
|\tilde{\varphi}_{l,m,n}^{\pm}\rangle=&\frac{1}{16\sqrt{2}}(
   \alpha_1|H_{l}\rangle_c|H_{m}\rangle_{t_1}|H_{n}\rangle_{t_2}
  \\&+\alpha_2|H_{l}\rangle_c|H_{m}\rangle_{t_1}|V_{n}\rangle_{t_2}
  \\&+\alpha_3|H_{l}\rangle_c|V_{m}\rangle_{t_1}|H_{n}\rangle_{t_2}
  \\&+\alpha_4|H_{l}\rangle_c|V_{m}\rangle_{t_1}|V_{n}\rangle_{t_2}
\\&\pm\alpha_5|V_{l}\rangle_c|H_{m}\rangle_{t_1}|H_{n}\rangle_{t_2}
\\&\pm\alpha_6|V_{l}\rangle_c|V_{m}\rangle_{t_1}|H_{n}\rangle_{t_2}
\\&\pm\alpha_7|V_{l}\rangle_c|H_{m}\rangle_{t_1}|V_{n}\rangle_{t_2}
\\&\pm\alpha_8|V_{l}\rangle_c|V_{m}\rangle_{t_1}|V_{n}\rangle_{t_2}).
\end{split}
\end{eqnarray}

Finally, the minus signs in Eq. (\ref{A29}) can be corrected by applying an HWP$^{0^\circ}$s  to mode 11 (12) of photon $c$ if the outgoing photons $c$, $t_1$, and $t_2$ are in modes 11, 11, and 12 or 11, 12, and 12 (12, 11, and 11 or 12, 12, and 11), respectively. Here, HWP$^{0^\circ}$ completes $|H\rangle\leftrightarrow|H\rangle$ and $|V\rangle\leftrightarrow-|V\rangle$.

Based on Eqs. (\ref{A1})-(\ref{A29}), one can see that Fig. \ref{post-selected-1} realizes a three-photon Fredkin gate with a success probability of $1/4\times1/4\times1/2=1/32$.

\vspace{6 cm}




\begin{thebibliography}{99}



\bibitem{book} M. A. Nielsen and I. L. Chuang, \emph{Quantum Computation and Quantum Information} (Cambridge University, Cambridge, 2000).


\bibitem{Barenco} A. Barenco, C. H. Bennett, R. Cleve, D. P. DiVincenzo, N. Margolus, P. Shor, T. Sleator, J. A. Smolin, and H. Weinfurter, Elementary gates for quantum computation, Phys. Rev. A \textbf{52}, 3457-3467 (1995).



\bibitem{ions1} C. Ospelkaus, U. Warring, Y. Colombe, K. R. Brown, J. M. Amini, D. Leibfried, and J. Wineland, Microwave quantum logic gates for trapped ions,  Nature (London) \textbf{476}, 181 (2011).

\bibitem{ions2} A. Bermudez, X. Xu, R. Nigmatullin, J. O'Gorman, V. Negnevitsky, P. Schindler, T. Monz, U. G. Poschinger, C. Hempel, J. Home, F. Schmidt-Kaler, M. Biercuk, R. Blatt, S. Benjamin, and M. M\"{u}ller, Assessing the progress of trapped-ion processors towards fault-tolerant quantum computation, Phys. Rev. X \textbf{7}, 041061 (2017).



\bibitem{NMR} G. R. Feng, G. F. Xu, and G. L. Long, Experimental realization of nonadiabatic holonomic quantum computation,  Phys. Rev. Lett. \textbf{110}, 190501 (2013).



\bibitem{Integrated-optics1}  A. Crespi, R.  Ramponi, R. Osellame, L.  Sansoni,  I.  Bongioanni, F. Sciarrino, G. Vallone, and P. Mataloni, Integrated photonic quantum gates for polarization qubits,  Nat. Commun. \textbf{2}, 566 (2011).

\bibitem{Integrated-optics2} J. Zeuner, A. N. Sharma, M. Tillmann, R. Heilmann, M. Gr\"{a}fe, A. Moqanaki, A. Szameit, and P. Walther, Integrated-optics heralded controlled-NOT gate for polarization-encoded qubits, npj Quantum Inform. \textbf{4}, 13 (2018).




\bibitem{atom-based1}B. Hacker, S. Welte, G. Rempe, and S. Ritter, A photon-photon quantum gate based on a single atom in an optical resonator, Nature (London) \textbf{536}, 7615 (2016).

\bibitem{atom-based2}D. Tiarks, S. Schmidt-Eberle, T. Stolz, G. Rempe, and S. D\"{u}rr,  A photon-photon quantum gate based on Rydberg interactions, Nat. Phys. \textbf{15}, 124 (2019).

\bibitem{multiphoton}S. Rosenblum, Y. Y. Gao, P. Reinhold, C. Wang, C. J. Axline, L. Frunzio, S. M. Girvin, L. Jiang, M. Mirrahimi, M. H. Devoret, and  R. J. Schoelkopf, A CNOT gate between multiphoton qubits encoded in two cavities, Nat. Commun. \textbf{9}, 652 (2018).

\bibitem{frequency} H. H. Lu, J. M. Lukens, B. P. Williams, P. Imany, N. A. Peters, A. M. Weiner, and P. Lougovski, A controlled-NOT gate for frequency-bin qubits, npj Quantum Inform. \textbf{5}, 24 (2019).


\bibitem{superconducting1} J. H. Plantenberg, P. C. de Groot, C. J. P. M. Harmans, and J. E. Mooij, Demonstration of controlled-NOT quantum gates on a pair of superconducting quantum bits, Nature (London) \textbf{447}, 836 (2007).

\bibitem{superconducting2} R. Barends, J. Kelly, A. Megrant, A. Veitia, D. Sank, E. Jeffrey, T. C. White, J. Mutus, A. G. Fowler, B. Campbell, Y. Chen, Z. Chen, B. Chiaro, A. Dunsworth, C. Neill, P. O’Malley, P. Roushan, A. Vainsencher, J. Wenner, A. N. Korotkov, A. N. Cleland, and J. M. Martinis, Superconducting quantum circuits at the surface code threshold for fault tolerance, Nature (London) \textbf{508}, 500 (2014).


\bibitem{hybrid} A. Reiserer, N. Kalb, G. Rempe, and S. Ritter, A quantum gate between a flying optical photon and a single trapped atom, Nature (London) \textbf{508}, 237 (2014).



\bibitem{atom1} S. Welte, B. Hacker, S. Daiss, S. Ritter, and G. Rempe, Photon-mediated quantum gate between two neutral atoms in an optical cavity, Phys. Rev. X \textbf{8}, 011018 (2018).

\bibitem{atom2} H. Levine, A. Keesling, G. Semeghini, A. Omran, T. T. Wang, S. Ebadi, H. Bernien,
M. Greiner, V. Vuleti\'{c}, H. Pichler, and M. D. Lukin, Parallel implementation of high-fidelity multiqubit gates with neutral atoms, Phys. Rev. Lett. \textbf{123}, 170503 (2019).



\bibitem{QD1} J. M. Nichol, L. A. Orona, S. P. Harvey, S. Fallahi, G. C. Gardner, M. J. Manfra, and A. Yacoby, High-fidelity entangling gate for double-quantum-dot spin qubits, npj Quantum Inform. \textbf{3}, 3 (2017).



\bibitem{NV} X. Rong, J. Geng, F. Shi, Y. Liu, K. Xu, W. Ma, F. Kong, Z. Jiang, Y. Wu, and J. Du, Experimental fault-tolerant universal quantum gates with solid-state spins under ambient conditions, Nat. Commun. \textbf{6}, 8748 (2015).






%
%
%
%
%
%
%
%
%








\bibitem{error-correction} M. D. Reed, L. DiCarlo, S. E. Nigg, L. Sun, L. Frunzio, S. M. Girvin, and R. J. Schoelkopf, Realization of three-qubit quantum error correction with superconducting circuits, Nature (London) \textbf{482}, 382 (2012).

\bibitem{Fault-tolerant} A. Paetznick and B. W. Reichardt, Universal fault-tolerant quantum computation with only transversal gates and error correction, Phys. Rev. Lett. \textbf{111}, 090505 (2013).

\bibitem{algorithms} C. Figgatt, D. Maslov, K. A. Landsman, N. M. Linke, S. Debnath, and C. Monroe, Complete 3-qubit Grover search on a programmable quantum computer, Nat. Commun.  \textbf{8},  1918 (2017).



\bibitem{entangling-operations} Y. Y. Gao, B. J. Lester, K. S. Chou, L. Frunzio, M. H. Devoret, L. Jiang, S. M. Girvin, and R. J. Schoelkopf, Entanglement of bosonic modes through an engineered exchange interaction,  Nature (London) \textbf{566}, 509 (2019).

\bibitem{Fredkin1} E. Fredkin and T. Toffoli, Conservative logic, Int. J. Theor. Phys. \textbf{21}, 219–253 (1982).

\bibitem{Fredkin2} T. Sleator and H. Weinfurter, Realizable universal quantum logic gates, Phys. Rev. Lett. \textbf{74}, 4087-4090 (1995).

\bibitem{Fredkin6}  H. F. Chau and F. Wilczek, Simple realization of the Fredkin gate using a series of two-body operators, Phys. Rev. Lett. \textbf{75}, 748-750 (1995).


\bibitem{Fredkin5-1} J. A. Smolin and D. P. DiVincenzo, Five two-bit quantum gates are sufficient to implement the quantum Fredkin gate, Phys. Rev. A \textbf{53}, 2855 (1996).


\bibitem{global} S. S. Ivanov, P. A. Ivanov, and N. V. Vitanov, Efficient construction of three- and four-qubit quantum gates by global entangling gates, Phys. Rev. A \textbf{91}, 032311 (2015).

\bibitem{Fredkin5-2} N. Yu  and M. Ying, Optimal simulation of Deutsch gates and the Fredkin gate, Phys. Rev. A \textbf{91}, 032302 (2015).


\bibitem{Fredkin-optimal1} V. V. Shende and I. L. Markov, On the CNOT-cost of TOFFOLI gates, Quantum Inf. Comput. \textbf{9}, 461-486 (2009).

\bibitem{Frekin-optimal2} T. Kim and B. S. Choi, Efficient decomposition methods for controlled-$R_n$ using a single ancillary qubit, Sci. Rep. \textbf{8}, 5445 (2018).


\bibitem{T-PRA} T. C. Ralph, K. J. Resch, and A. Gilchrist, Efficient Toffoli gates using qudits, Phys. Rev. A \textbf{75}, 022313 (2007).

\bibitem{T-NatPhy} B. P. Lanyon, M. Barbieri, M. P. Almeida, T. Jennewein, T. C. Ralph, K. J. Resch, G. J. Pryde, J. L. O'Brien, A. Gilchrist, and  A. G. White, Simplifying quantum logic using higher-dimensional Hilbert spaces, Nat. Phys. \textbf{5}, 134 (2009).

\bibitem{multilevel4} R. Ionicioiu, T. P. Spiller, and W. J. Munro, Generalized Toffoli gates using qudit catalysis, Phys. Rev. A \textbf{80}, 012312 (2009).

\bibitem{Multivalue} W. D. Li, Y. J. Gu, K. Liu, Y. H. Lee, and Y. Z. Zhang, Efficient universal quantum computation with auxiliary
Hilbert space, Phys. Rev. A \textbf{88}, 034303 (2013).

\bibitem{lower-T} N. Yu, R. Duan, and M. Ying, Five two-qubit gates are necessary for implementing the Toffoli gate, Phys. Rev. A \textbf{88}, 010304(R) (2013).


\bibitem{Liu} W. Q. Liu and H. R. Wei, Optimal synthesis of the Fredkin gate in a multilevel system, New J. Phys. \textbf{22}, 063026 (2020).




\bibitem{construction1} J. Fiur\'{a}\v{s}ek, Linear-optics quantum Toffoli and Fredkin gates, Phys. Rev. A \textbf{73}, 062313 (2006).

\bibitem{Gong} Y. X. Gong, G. C. Guo, and T. C. Ralph, Methods for a linear optical quantum Fredkin gate, Phys. Rev. A \textbf{78}, 012305 (2008).

\bibitem{construction2} J. Fiur\'{a}\v{s}ek,  Linear optical Fredkin gate based on partial-SWAP gate,  Phys. Rev. A \textbf{78}, 032317 (2008).

\bibitem{Ono} T. Ono, R. Okamoto, M. Tanida, H. F. Hofmann, and S. Takeuchi, Implementation of a quantum controlled-SWAP gate with photonic circuits,  Sci. Rep. \textbf{7}, 45353 (2017).



\bibitem{Patel} R. B. Patel, J. Ho, F. Ferreyrol, T. C. Ralph, and G. J. Pryde,  A quantum Fredkin gate, Sci. Adv. \textbf{2}, e1501531 (2016).


\bibitem{polarization-saptial} R. St\'{a}rek, M. Mi\v{c}uda, M. Mikov\'{a}, I. Straka, M. Du\v{s}ek, P. Marek, M. Je\v{z}ek, R. Filip, and J. Fiur\'{a}\v{s}ek, Nondestructive detector for exchange symmetry of photonic qubits,  npj Quantum Inform. \textbf{4}, 35 (2018).


\bibitem{polarization-OAM} D. F. Urrego, D. Lopez-Mago, V. Vicu\~{n}a-Hern\'{a}ndez, and J. P. Torres, Quantum-inspired Fredkin gate based on spatial modes of light, Opt. Express  \textbf{28}, 12661 (2020).







\bibitem{CNOT-BS1} H. F. Hofmann and S. Takeuchi, Quantum phase gate for photonic qubits using only beam splitters and postselection, Phys. Rev. A \textbf{66}, 024308 (2002).

\bibitem{CNOT-BS2} J. L. O’Brien, G. J. Pryde,  A. G. White,  T. C. Ralph,  and D. Branning, Demonstration of an all-optical quantum
controlled-NOT gate, Nature (London) \textbf{426}, 264 (2003).

\bibitem{CNOT-BS3} N. K. Langford, T. J. Weinhold, R. Prevedel, K. J. Resch, A. Gilchrist, J. L. O’Brien, G. J. Pryde, and A. G. White, Demonstration of a simple entangling optical gate and its use in Bell-state analysis,  Phys. Rev. Lett. \textbf{95}, 210504 (2005).

\bibitem{CNOT-BS4} N. Kiesel, C. Schmid, U. Weber, R. Ursin, and H. Weinfurter, Linear optics controlled-phase gate made simple, Phys. Rev. Lett. \textbf{95}, 210505 (2005).

\bibitem{opti-CNOT}  T. B. Pittman, B. C. Jacobs, and J. D. Franson, Probabilistic quantum logic operations using polarizing beam
splitters, Phys. Rev. A \textbf{64}, 062311 (2001).

\bibitem{Shende} V. V. Shende, I. L. Markov, and S. S. Bullock, Minimal universal two-qubit controlled-NOT-based circuits, Phys. Rev. A \textbf{69}, 062321 (2004).


\bibitem{F-superconduct} A. Fedorov, L. Steffen, M. Baur, M. P. da Silva, and A. Wallraff, Implementation of a Toffoli gate with superconducting circuits, Nature (London) \textbf{481}, 170 (2012).


\bibitem{KLM}  E. Knill,  R. Laamme, and G. J. Milburn, A scheme for efficient quantum computation with linear optics, Nature (London) \textbf{409}, 46-52 (2001).

\bibitem{LOQC} P. Kok, W. J. Munro, T. C. Ralph, J. P. Dowling, and G. J. Milburn, Linear optical quantum computing with photonic qubits, Rev. Mod. Phys. \textbf{79}, 135-174 (2007).

\bibitem{Loss-tolerant}  T. C. Ralph, A. J. F. Hayes, and A. Gilchrist, Loss-tolerant optical qubits, Phys. Rev. Lett. \textbf{95},  100501 (2005).





\end{thebibliography}
\end{document}